\documentclass[reprint,amsmath,amssymb,aps, superscriptaddress, nofootinbib]{revtex4-1}

\usepackage{graphicx}%
\usepackage{multirow}
\usepackage{comment}
\usepackage{bm}
\usepackage[colorlinks=true,citecolor=blue,linkcolor=blue]{hyperref}
 \renewcommand\arraystretch{1.3}

\def\eqref#1{Eq.~(\ref{#1})}
\def\Fig#1{Fig.~\ref{#1}}
\def\st{\begin{equation}}
\def\stp{\end{equation}}
\usepackage{array}   %
\newcolumntype{L}{>{$}l<{$}} %
\def\app#1{Appendix~\ref{#1}}
\def\k{\bm k}
\def\kb{{k'}}
\def\pb{{p'}}
\def\M{\mathcal M}
\def \cf {C_F}

\def \ca {C_A}

\newcommand{\pcut}{p_{\textrm{cut}}}
\newcommand{\mutildeqperp}{\mu_{\tilde{q}_\perp}}
\newcommand{\nn}{\nonumber}

\begin{document}
\title{Parton energy loss in a hard-soft factorized approach}

\author{Tianyu Dai}
\email{td115@duke.edu}
\affiliation{Department of Physics, Duke University, Durham, North Carolina 27708-0305, USA}
\author{Jean-Fran\c{c}ois Paquet}
\affiliation{Department of Physics, Duke University, Durham, North Carolina 27708-0305, USA}
\author{Derek Teaney}
\affiliation{Department of Physics and Astronomy, Stony Brook University, New York 11794-3800, USA}
\author{Steffen A. Bass}
\affiliation{Department of Physics, Duke University, Durham, North Carolina 27708-0305, USA}

\date{\today}
\begin{abstract}
An energetic parton travelling through a quark-gluon plasma loses energy via occasional hard scatterings and frequent softer interactions.
Whether or not these interactions admit a perturbative description, the effect of the soft interactions can be factorized and encoded in a small number of transport coefficients.
In this work, we present the numerical implementation of a hard-soft factorized parton energy loss model which combines a stochastic description of soft interactions and rate-based modelling of hard scatterings.
We introduce a scale to estimate the regime of validity of the stochastic description, allowing for a better understanding of the model's applicability at small and large coupling. 
We study the energy and fermion-number cascade of energetic partons as an application of the model.
\end{abstract}

\maketitle
\section{Introduction}

The production of energetic hadrons and jets in heavy ion collisions is markedly different from the production of energetic electroweak bosons. The latter clearly exhibit ``binary scaling'': weak bosons and high-energy photons are produced as if nucleons from each nucleus were independently undergoing inelastic binary collisions~\cite{CMS:2012fgk,CMS:2012oiv,ATLAS:2012qdj,ATLAS:2014sic,ATLAS:2015rlt,CMS:2014dyj,PHENIX:2005yls} (see also Refs.~\cite{Miller:2007ri,Armesto:2015ioy} and references therein). 
Hadron and jet measurements, on the other hand, display evident deviations from binary scaling.
These deviations are understood to be
a consequence of the formation of a quark-gluon plasma in relativistic nuclear collisions: energetic parton \emph{production} does follow ``binary scaling''; it is their subsequent interactions with the plasma that lead to parton energy loss, and consequently to an apparent deviation from binary scaling for hadronic observables.

This characteristic phenomena of jet and hadron ``energy loss'' in heavy ion collisions has been observed at both the Relativistic Heavy-Ion Collider (RHIC) and the Large Hadron Collider (LHC)~\cite{Majumder:2010qh, Muller:2012zq, Mehtar-Tani:2013pia,Roland:2014jsa, Connors:2017ptx, Busza:2018rrf}. 
Energetic partons are produced at the earliest stage of heavy ion collisions, and they propagate through all the different phases of the collisions. 
As a consequence of their interactions
with the quark-gluon plasma, the momentum distribution of these energetic partons changes distinctly compared to the baseline observed in proton-proton collisions.\footnote{The role of hadronic energy loss is still under investigation. See Ref.~\cite{Dorau:2019ozd} for example.}
This makes them important probes of the deconfined nuclear plasma produced in heavy ion collisions.

A number of different formalisms have been used to model the interaction of energetic light partons\footnote{We use a parameter ``$p_{cut}$'' to define which partons we consider as ``energetic''. We only track the propagation of these energetic partons with $p>p_{cut}$. We use $p_{cut}= 2$~GeV throughout this work since we focus on the energy loss of light partons.} with the constituents of the plasma~\cite{Baier:1996kr, Baier:1996sk,Zakharov:1996fv,Zakharov:1997uu,Gyulassy:1999zd,Gyulassy:2000fs,Gyulassy:2000er,Guo:2000nz,Arnold:2002zm, Jeon:2003gi, Armesto:2003jh} (see also Refs.~\cite{Bass:2008rv,Majumder:2010qh,CaronHuot:2010bp,Armesto:2011ht} and references therein).
Fundamentally, most parton energy loss formalisms have a well-understood common core, yet applications to heavy ion collisions tend to require approximations and practical considerations that lead to non-negligible differences between parton energy loss models~\cite{Bass:2008rv,CaronHuot:2010bp,Armesto:2011ht}.
One difference between the models is the treatment of the underlying plasma, which %
is often assumed
to be made of a large number of quarks and gluons with energies of $\lesssim 1$~GeV  in near local thermal equilibrium.
Whether these quarks and gluons are treated as dynamical entities or as static scattering centers is one of many differences in the energy loss formalisms~\cite{Bass:2008rv,CaronHuot:2010bp,Armesto:2011ht}.
The above assumption is important, given that  the quark-gluon plasma produced in heavy ion collisions is understood to be strongly coupled~\cite{Shuryak:2014zxa}, and a  quasiparticle description may not be justified.\footnote{In particular,  hydrodynamic simulations of this plasma's evolution do not rely on a quasiparticle picture of deconfined nuclear matter until hadronization.}

A different phrasing of the above challenge is that the energy loss of even very energetic partons can be affected by non-perturbative effects from the strongly-coupled plasma.
Hard interactions --- those with large momentum transfer between the energetic parton and the plasma --- are expected to have smaller non-perturbative effects, or even be accessible perturbatively, as a consequence of the running of the QCD coupling.
On the other hand, ``soft'' parton-plasma interactions with small momentum transfer are expected to suffer the largest non-perturbative effects.\footnote{Note other works such as Refs.~\cite{Casalderrey-Solana:2014bpa,Casalderrey-Solana:2016jvj} assume that neither soft or hard interactions are perturbative, and consequently evaluate parton energy loss using gauge-field duality.} We note that ``hard interactions'' and ``soft interactions'' have various meanings in the literature, but for the purpose of this work, the temperature of the plasma can be considered as the scale separating hard (larger than $T$) and soft (smaller than $T$) interactions.

A stochastic treatment of these soft interactions of energetic partons provides an alternative approach to account for non-perturbative effects --- an approach that is agnostic to the strongly- or weakly-coupled nature of the underlying deconfined plasma.
The dynamical details of the large number of soft interactions are encoded in a small number of transport coefficients.
The latter can be parametrized and constrained from measurements. They can also be studied using lattice techniques~(see for example Ref.~\cite[Section 4]{Ghiglieri:2015zma} and Refs.~\cite{Moore:2019lgw,Schlusser:2020lme} and references therein). 
From a practical point of view, a stochastic description of a large number of soft interactions can also be more efficient numerically than a rate-based approach. 

A systematic hard-soft factorization of parton energy loss was proposed recently to describe parton propagation in a weakly-coupled QGP~\cite{Ghiglieri:2015ala,Ghiglieri:2015zma}. In this factorization, soft interactions are described as a stochastic process with drag and diffusion transport coefficients calculated perturbatively; hard interactions are solved with rates that are also calculated perturbatively. 
In the weakly-coupled regime, parton energy loss in this hard-soft factorizated scheme was shown to be equivalent to a fully rate-based treatment of parton energy loss~\cite{Ghiglieri:2015ala}. 
Importantly parton energy loss in this hard-soft factorization can also be extended to next-to-leading order~\cite{Ghiglieri:2015ala}, a feature beyond the scope of this work which we shall explore in the future. 

As discussed above, the drag and diffusion contribution to parton energy loss can be factorized systematically, and calculated non-perturbatively e.g. based on Electrostatic Quantum Chromodynamics (EQCD)~\cite{Moore:2019lgw}, or fitted to data.
These extractions will then depend on the separation scale $\mu$, which appears in the approach. At higher order,  the drag and diffusion coefficients will 
evolve with the scale $\mu \sim \pi T$, incorporating in a consistent way the running of the coupling. While this is beyond the scope of this work,
we hope that this manuscript can provide a first step in that direction. Throughout the paper we will already study the dependence of various observables
on the separation scale $\mu$,  and, encouragingly, find that this dependence is moderate in most cases.

The above work is based on the ``effective kinetic theory'' approach~\cite{Arnold:2002zm} derived for a weakly-coupled quark-gluon plasma.
In a weakly-coupled plasma, quark and gluon excitations are described as quasi-particles with effective properties related to the local density of the plasma. In this effective kinetic approach, the dynamics of quasi-particles are described by Boltzmann transport equations.
Leading order [$\mathcal{O}(\alpha_s)$] realizations of this effective kinetic approach --- extrapolated to large values of strong coupling constant $\alpha_s$ --- have been used widely to study parton energy loss~(see e.g. Refs.~\cite{Jeon:2003gi,Qin:2007rn, Schenke:2009gb,Burke:2013yra, Cao:2017zih}). 

In this work, we present the first numerical implementation of the hard-soft factorized parton energy loss model~\cite{Ghiglieri:2015ala} discussed above. 
For our implementation we utilize the publicly available JETSCAPE framework~\cite{Putschke:2019yrg}, as it allows us a straightforward integration of our parton energy loss model with the other ingredients necessary for a full simulation of jet production in heavy ion collisions.
We first test and validate this factorization of parton energy loss in the weak coupling regime for a static medium.

We introduce a dimensionless scale to quantify the kinematic range for which soft interactions can be described accurately with a stochastic approach.
We use this scale to discuss a hard-soft factorization model for a strongly-coupled quark-gluon plasma, relevant for phenomenological applications in heavy ion collisions.

Finally, we present an application of our new factorized model of parton energy loss by calculating the energy and fermion-number cascade of an energetic parton propagating in a static medium, finding good agreement with known analytical approximations. 

\section{Hard-soft factorization of parton energy loss in the weakly-coupled regime: theory}
\subsection{Effective kinetic approach in weakly-coupled regime}

The evolution of an energetic parton in a thermal medium of temperature T can be described by a Boltzmann transport equation~\cite{Arnold:2003zc}:
\begin{equation}
    \label{eq:Boltzmann}
    \left( \frac{\partial}{\partial t}+ \frac{\textbf{p}}{p} \cdot \mathcal{\nabla} \right) \delta f_a = - \mathcal{C}[\delta f_a, n_a]
\end{equation}
where $P=(p,\textbf{p})$ is the four-momentum of the energetic parton 
and $\mathcal{C}$ is the collision kernel of the parton with the medium. The index $a$ represents partons with a certain color and helicity state. 
We use the same notation for the parton momentum distributions as in Ref.~\cite{Ghiglieri:2015ala}:  the distribution of rare energetic partons of type $a$ is $\delta f^a(\mathbf{p}, \mathbf{x}, t)$, to distinguish it from the quasi-thermal distribution of soft particles $n^a(\mathbf{p}, T(\mathbf{x}, t), \mathbf{u}(\mathbf{x}, t))$, where $\mathbf{u}$ is the flow velocity. In this notation, the total phase space distribution of quasiparticle $a$ is $f^a(\mathbf{p}, \mathbf{x}, T) = n^a(\mathbf{p}, T(\mathbf{x}, t), \mathbf{u}(\mathbf{x}, t))+\delta f^a(\mathbf{p}, \mathbf{x}, t)$.
We assume $p \gg T$ and $g \equiv \sqrt{\frac{\alpha_s}{4\pi}} \ll 1$. 
Because interactions between energetic partons themselves are rare and can be neglected, the Boltzmann equation is effectively linear in $\delta f_a$. 

At leading order, the interactions between quasi-particles can be divided as $2\leftrightarrow 2$ elastic interactions and $1\leftrightarrow 2$ inelastic interactions. Elastic $2\leftrightarrow 2$ processes refer to elementary scatterings involving two incoming particles and two outgoing particles without any radiation. Multiple soft $2\leftrightarrow 2$ scatterings between the energetic parton and the plasma can induce a collinear radiation. In the effective kinetic approach, these multiple soft scatterings are resummed consistently, to account for interference between subsequent collisions which lead to the Landau-Pomeranchuk-Migdal (LPM) effect.
This resummed collinear radiation is known as the effective $1\leftrightarrow 2$ process. 
The collision kernel of both $2\leftrightarrow 2$ and $1\leftrightarrow 2$ processes can be written as 
\begin{equation}
    \mathcal{C} = \mathcal{C}^{1\leftrightarrow 2}+\mathcal{C}^{2\leftrightarrow 2}. 
\end{equation}

Importantly, in our approach, we only follow the evolution of energetic partons with an energy above a cutoff $\pcut = 2$~GeV. 
Our assumption is that we should focus our efforts on high-$p_T$ observables which are dominated by partons above this cutoff.
After neglecting terms suppressed by $\exp(-p/T)$, the collision kernels $\mathcal{C}^{1\leftrightarrow 2}$ and $\mathcal{C}^{2\leftrightarrow 2}$ read~\cite{Ghiglieri:2015ala}: 
\begin{widetext}
\begin{equation}
\label{eq:1to2}
\begin{split}
    \mathcal{C}_a^{1\leftrightarrow 2}\left[\delta f\right] &= \frac{(2\pi)^3}{2|\bm{p}|^2\nu_a}\sum_{bc}\int^\infty_0dp'dq'\gamma_{bc}^{a}(\bm{p}; p'\hat{\bm{p}}, q'\hat{\bm{p}})\delta(|\bm{p}|-p'-q')\\
    &\times \left\{\delta f^a(\bm{p})\left[1\pm n^b(p')\pm n^c(q')\right]-\left[\delta f^b(p'\hat{\bm{p}})n^c(q')+n^b(p')\delta f^c(k'\hat{\bm{p}})\right]\right\}\\
    &+\frac{(2\pi)^3}{|\bm{p}|^2\nu_a}\sum_{bc}\int^\infty_0dqdp'\gamma_{ab}^{c}(p'
    \hat{\bm{p}}; \bm{p}, q\hat{\bm{p}})\delta(|\bm{p}|+q-p')\\
    &\times \left\{\delta f^a(\bm{p})n^b(q)-\delta f^c(p'\hat{\bm{p}})\left[1\pm n^b(q)\right]\right\}\, , 
\end{split}
\end{equation}

\begin{equation}
\label{eq:2to2}
\begin{split}
    \mathcal{C}_a^{2\leftrightarrow 2}\left[\delta f\right] =& \frac{1}{4|\bm{p}|\nu_a}\sum_{bcd}\int_{\bm{k}\bm{p}'\bm{k}'}|\mathcal{M}_{cd}^{ab}(\bm{p}, \bm{k}; \bm{p}', \bm{k}')|^2(2\pi)^4\delta^{(4)}(P+K-P'-K')\\
    &\times \left\{\delta f^a(\bm{p})n^b(k)\left[1\pm n^c(p')\pm n^d(k')\right]-\delta f^c(\bm{p}')n^d(k')\left[1\pm n^b(k)\right]-n^c(p')\delta f^d(\bm{k}')\left[1\pm n^b(k)\right]\right\}\, , 
\end{split}
\end{equation}
\end{widetext}
where the notation for the Lorentz-invariant integration is
\begin{equation}
    \int_{\bm{k}} \dots \equiv \int \frac{d^3k}{2k(2\pi)^3}\dots
\end{equation}
and $\nu_a$ is the degeneracy of particle $a$. 

For $\mathcal{C}^{1\leftrightarrow 2}$, $a$ is the incoming hard parton with the momentum $\bm{p}$, and $b, c$ are outgoing particles with the momentum $\bm{p}', \bm{k}'$. $\gamma_{bc}^{a}$ is the splitting kernel of $a\rightarrow bc$, which can be calculated with the AMY integral equations~\cite{Arnold:2002zm,Jeon:2003gi}. 

\begin{figure}
    \centering
    \includegraphics[width=\columnwidth]{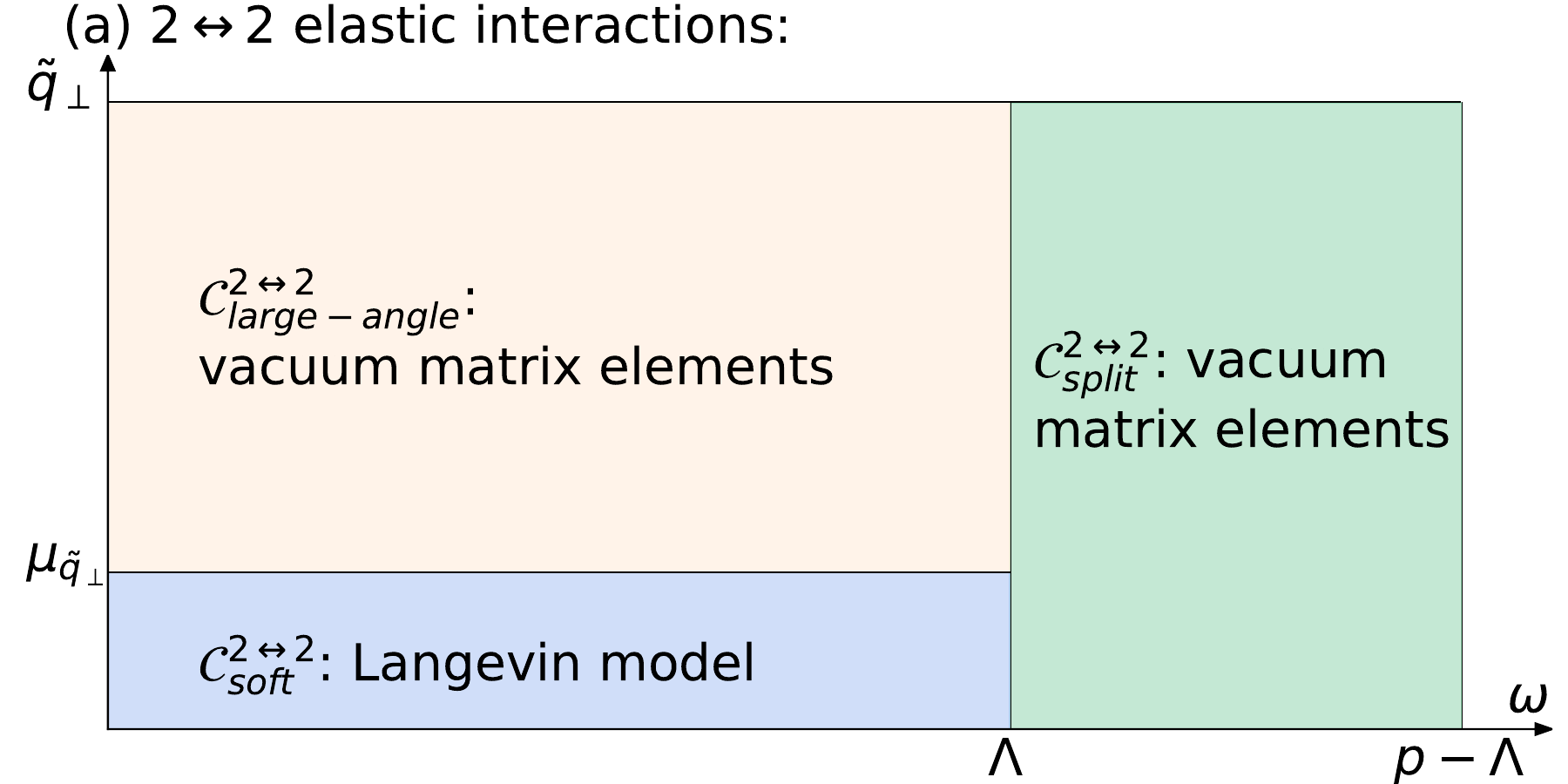}
    \includegraphics[width=\columnwidth]{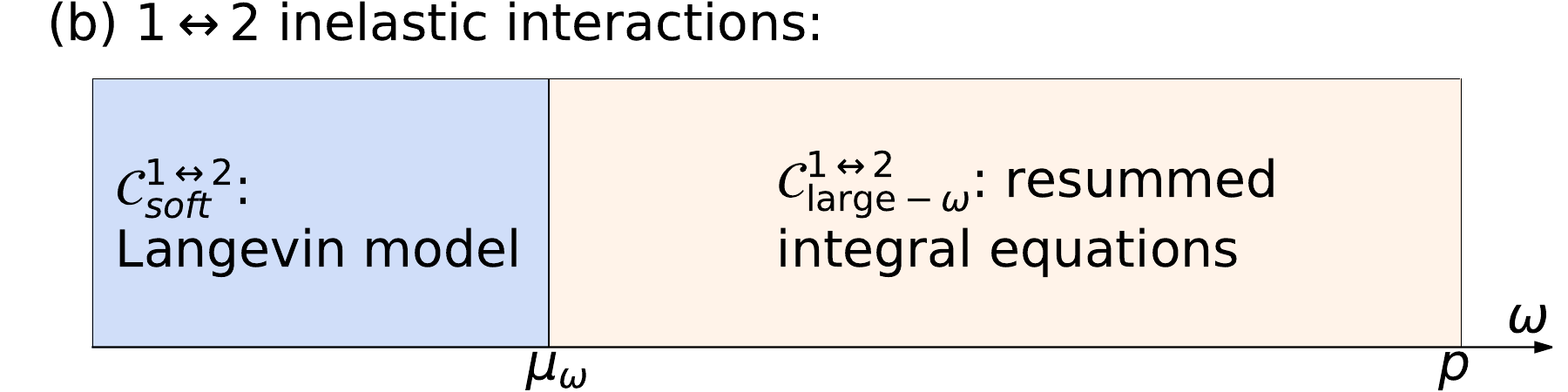}
    \caption{Treatment of different processes in the hard-soft factorized parton energy loss model}
    \label{fig:reformulation}

\end{figure}

For $\mathcal{C}^{2\leftrightarrow 2}$, particle $a$ is the incoming energetic parton with momentum $\bm{p}$, particle $b$ is the plasma particle with the momentum $\bm{k}$ interacting with $a$, and particle $c, d$ are the outgoing particles with momentum $\bm{p}', \bm{k}'$. $\mathcal{M}_{cd}^{ab}$ is the matrix element of the elementary process $ab\rightarrow cd$~\cite{Arnold:2002zm}. 

\subsection{Reformulating parton energy loss with hard-soft factorization}
\label{sec:factorized_model}

In the hard-soft factorized parton energy loss model introduced in Ref.~\cite{Ghiglieri:2015ala}, $1\leftrightarrow 2$ and $2\leftrightarrow 2$ processes are further divided into soft interactions and hard interactions. The collision kernel is rewritten as 
\begin{equation}
\mathcal{C} = \mathcal{C}^{1\leftrightarrow 2}_{\textrm{hard}}+\mathcal{C}^{1\leftrightarrow 2}_{\textrm{soft}}+\mathcal{C}^{2\leftrightarrow 2}_{\textrm{hard}}+\mathcal{C}^{2\leftrightarrow 2}_{\textrm{soft}}\, . 
\end{equation}

In this hard-soft factorized model, soft interactions described by $\mathcal{C}^{1\leftrightarrow 2}_{\textrm{soft}}$ and $\mathcal{C}^{2\leftrightarrow 2}_{\textrm{soft}}$ are treated stochastically with the Langevin equation. 

Hard inelastic interactions,  $\mathcal{C}^{1\leftrightarrow 2}_{\textrm{hard}}$, are treated with an emission rate as calculated from the AMY integral equations~\cite{Arnold:2002zm}.\footnote{We thank Guy D. Moore for his numerical solver for AMY integral equations.} We refer to them
as large-$\omega$ interactions.

The hard $2\leftrightarrow 2$ part $\mathcal{C}^{2\leftrightarrow 2}_{\textrm{hard}}$ is further divided as (i) large-angle interactions, and (ii) splitting approximation processes, based on the energy transfer $\omega$:  
\begin{equation}
\mathcal{C}^{2\leftrightarrow 2}_{\textrm{hard}} = \mathcal{C}_{\textrm{large-angle}}^{2\leftrightarrow2}+\mathcal{C}_{\textrm{split}}^{2\leftrightarrow2}\, . 
\end{equation}
The physical meaning of this separation is the following. Elastic collisions occur between an energetic parton ($p\gg T$) and a lower energy quasi-thermal quark or gluon ($k \sim T$). On rare occasions, the momentum transfer in these elastic collisions is sufficient to make the low-energy quark or gluon become an energetic parton with $k\gg T$; such partons are referred to in the literature as ``recoil partons''. The process through which a recoil parton is produced  is akin to a splitting process: a single energetic particle splits in two energetic ones. The kinematic of this process simplifies and it benefits from being treated separately. %

The factorization of the phase space for this reformulation is summarized in Fig. \ref{fig:reformulation}. An ensemble of cutoffs is used to divide the different regions of phase space. We discuss the details of the different treatments and these cutoffs in the following subsections. 

\subsubsection{Treatment of hard interactions: inelastic case ($1\leftrightarrow 2$)}

\label{sec:factorized_model:inelastic_hard}

\begin{figure}
    \centering
    \includegraphics[width=0.8\linewidth]{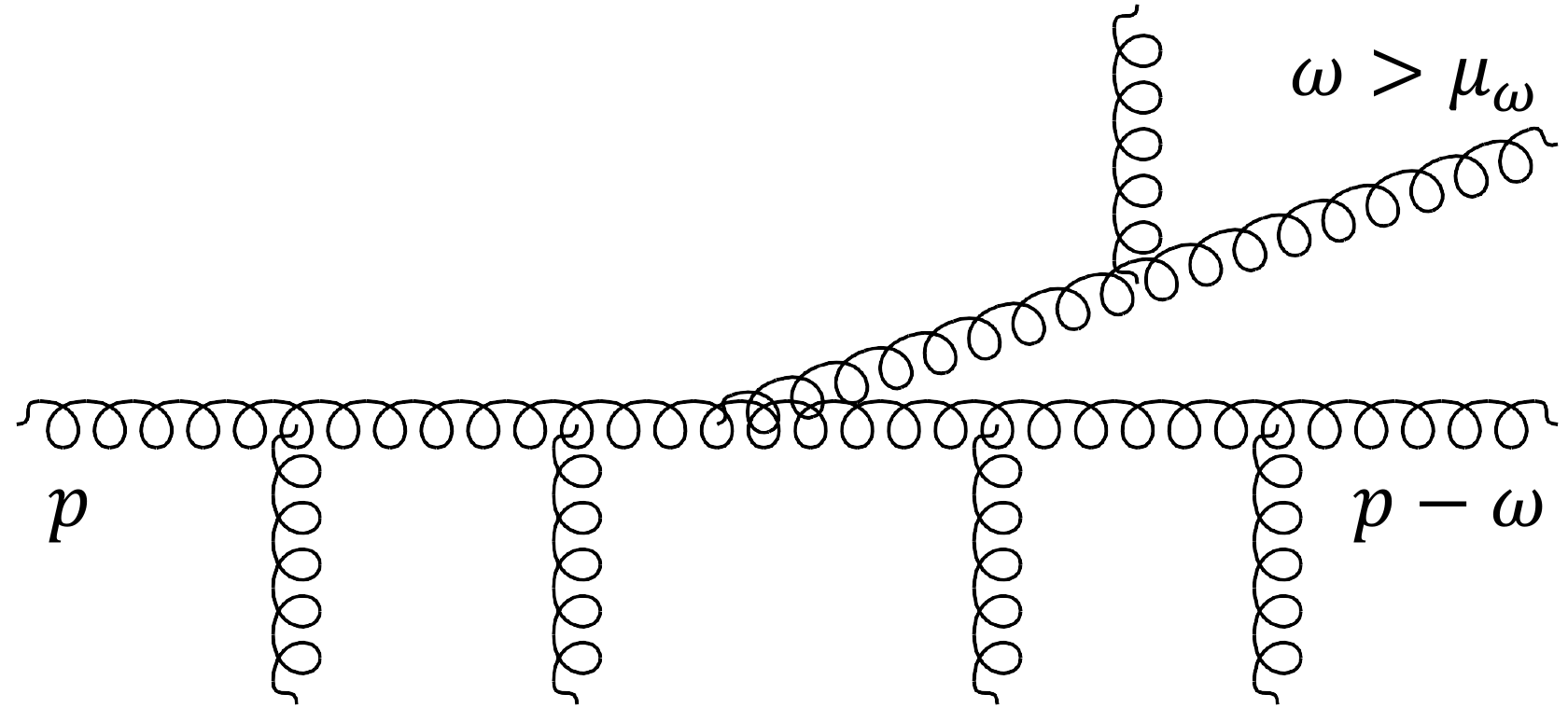}
    \caption{Example of inelastic interaction, in which multiple soft scatterings induce the radiation of a soft gluon with energy $\omega$. We denote the radiations with $\omega > \mu_\omega$ as large-$\omega$ inelastic interactions.
    }
    \label{fig:inel_diagram}
\end{figure}

The diagram of a typical $1\leftrightarrow 2$ inelastic interaction is shown as Fig. \ref{fig:inel_diagram}. We assume the energy of the radiated particle is $\omega$. We define a hard-soft cutoff $\mu_\omega$ based on the radiated energy $\omega$, to divide $\mathcal{C}^{1\leftrightarrow 2}_{\textrm{hard}}$ and $\mathcal{C}^{1\leftrightarrow 2}_{\textrm{soft}}$. 
In the weakly-coupled regime, the cutoff $\mu_\omega$ is limited to $\mu_\omega \lesssim T$, where $T$ is the temperature of the thermal medium. 

Collinear radiations with energy $\omega  > \mu_\omega $ are included into the hard part, $\mathcal{C}^{1\leftrightarrow 2}_{\textrm{hard}}$; 
they are treated as usual with emission rates calculated from AMY's integral equations as in Eq. (\ref{eq:1to2}). 

\subsubsection{Treatment of hard interactions: elastic case ($2\leftrightarrow 2$)}

\label{sec:factorized_model:elastic_hard}

\begin{figure}
    \centering
    \includegraphics[width=0.45\linewidth]{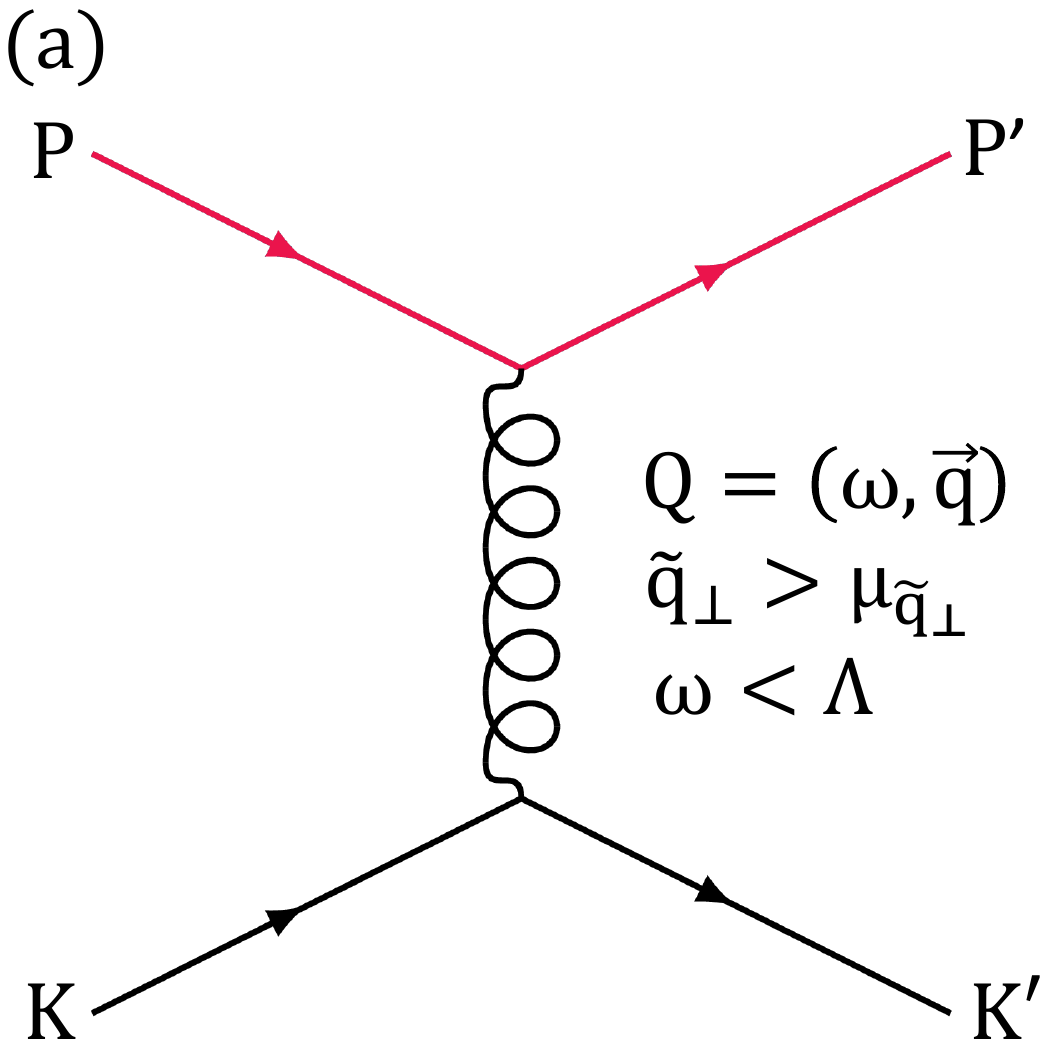}
    \includegraphics[width=0.45\linewidth]{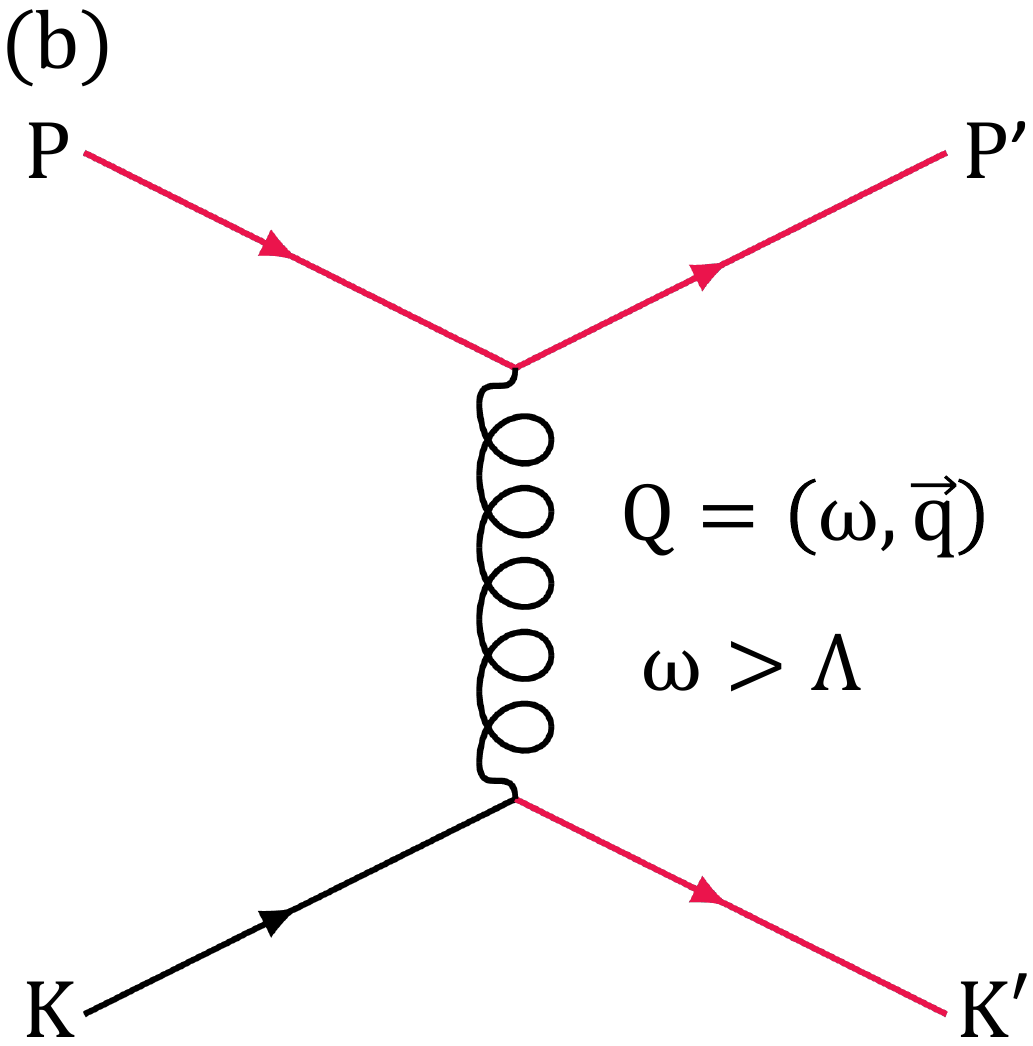}
    \caption{(a) Example of large-angle elastic $2\leftrightarrow 2$ interactions, where $\tilde{q}_\perp > \mu_{\tilde{q}_\perp}$ and $\omega < \Lambda$; (b) example of elastic $2\leftrightarrow 2$ interactions with $\omega > \Lambda$, which is treated with a splitting approximation (see text). 
    }
    \label{fig:elas_diagram}
\end{figure}
The diagram of a typical $2\leftrightarrow 2$ elastic interaction is shown in Fig. \ref{fig:elas_diagram}. We define the momentum transfer between the two incoming particles as $Q=(\omega, \vec{q})$; the four-momenta of the incoming and outgoing particles are $P=(p, \mathbf{p})$, $K=(k, \mathbf{p})$, $P' = P-Q$, and $K'=K+Q$. Using $\tilde{q}_\perp \equiv \sqrt{q^2-\omega^2}$ and $\omega$,  we divide the phase space of elastic interactions as
\begin{itemize}
    \item Large-angle scattering $\mathcal{C}_{\textrm{large-angle}}^{2\leftrightarrow2}$:  $\tilde{q}_\perp > \mu_{\tilde{q}_\perp}$ and $\omega < \Lambda$;
    \item ``Splitting-like'' process  $\mathcal{C}_{\textrm{split}}^{2\leftrightarrow2}$: $\omega > \Lambda$
\end{itemize}
with
\begin{equation}
\mathcal{C}^{2\leftrightarrow 2}_{\textrm{hard}} = \mathcal{C}_{\textrm{large-angle}}^{2\leftrightarrow2}+\mathcal{C}_{\textrm{split}}^{2\leftrightarrow2}.
\end{equation}

\paragraph{Large-angle scattering ($\mathcal{C}_{\textrm{large-angle}}^{2\leftrightarrow2}$)}

Hard scatterings with $\tilde{q}_\perp > \mu_{\tilde{q}_\perp}$ and $\omega < \Lambda$ are denoted as large-angle interactions, because the scattering angle 
\begin{equation}
\frac{q_\perp}{q_z} \approx \frac{\tilde{q}_\perp}{\omega} 
\end{equation}
is generally large in this region. The cutoff $\mu_{\tilde{q}_\perp}$ is typically assumed to be $gT \ll \mu_{\tilde{q}_\perp} \ll T$~\cite{Ghiglieri:2015ala}, although we will see in Section~\ref{sec:tequila_weak_coupling} that this condition can be relaxed. We assume $p \gg \omega$ in this region, and simplify the matrix elements accordingly. 

We use vacuum matrix elements for $\mathcal{C}^{2\leftrightarrow 2}_{\textrm{hard}}$, because the screening effects are only significant for soft interactions ($\mathcal{C}^{2\leftrightarrow 2}_{\textrm{soft}}$) in the weakly-coupled regime~\cite{Ghiglieri:2015ala}. Since we are only interested in the evolution of energetic partons, we keep terms to the first order in $T/p$ in the matrix elements. 

The treatment of the region $\tilde{q}_\perp > \mu_{\tilde{q}_\perp}$ and $p-\Lambda < \omega < p$ --- which is handled differently for technical reasons --- is discussed in \app{app:remaining}. 

\paragraph{Splitting approximation ($\mathcal{C}_{\textrm{split}}^{2\leftrightarrow2}$)}
When both of the outgoing particles of a $2\leftrightarrow 2$ interaction are hard ($p', k' > \pcut$), the interaction can be effectively considered as a splitting process. The splitting leads to a hard recoil parton which should be included in the calculation. 

We use a cutoff $\Lambda$ on $\omega$ to distinguish two hard outgoing particles from only one hard outgoing particles. In principle, this cutoff $\Lambda$ should be $3T \ll \Lambda \ll p$. In the numerical implementation, unless specified otherwise, we choose $\Lambda = \min(\sqrt{3pT}, \pcut)$ to divide $\mathcal{C}^{2\leftrightarrow 2}_{\mathrm{split}}$ and $\mathcal{C}^{2\leftrightarrow 2}_{\mathrm{large-angle}}$. Recall that we use $\pcut=2$~GeV in this work. As shown in Fig. \ref{fig:elas_diagram}, splitting approximation process is the $2\leftrightarrow 2$ interactions with $\Lambda < \omega < p-\Lambda$.  

\begin{figure}
    \centering
    \includegraphics[width=\linewidth]{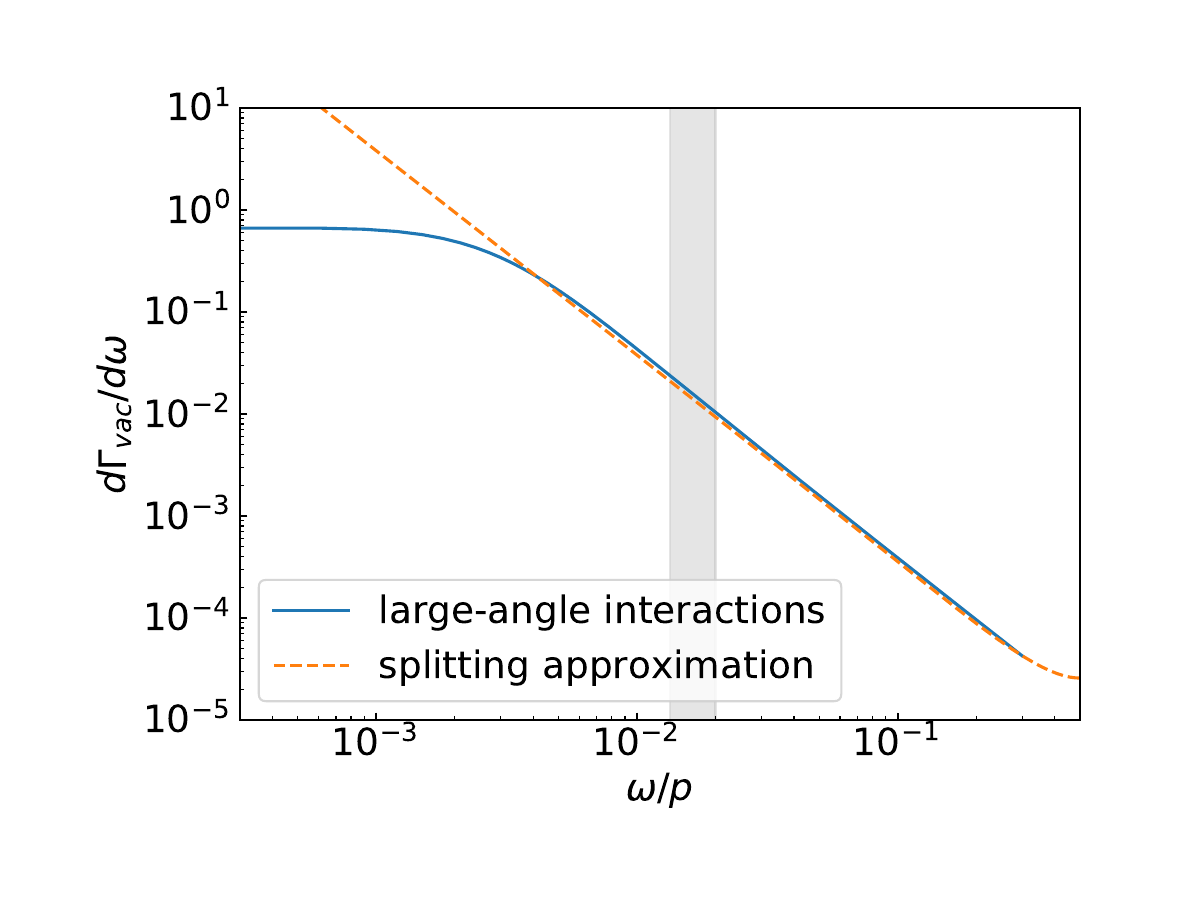}
    \caption{The differential rate of splitting approximation interactions and large-angle interactions for $gg \leftrightarrow gg$ process when $\alpha_s=0.3$. The shaded area is the region of $\sqrt{3p_{\textrm{cut}}T} < \omega < p_{\textrm{cut}}$. $d\Gamma_{\textrm{vac}}/d\omega$ is the differential rate of vacuum matrix elements for $2\leftrightarrow 2$ interactions. The results are for $p_0=100$~GeV. Note that in the numerical implementation, we double-count the large-angle interaction rate because we only sample in half of the phase space. Here, to compare with splitting approximation rate, we decrease the large-angle interaction rate in the numerical implementation by a factor of $\frac{1}{2}$ to cancel out the double-count. }
    \label{fig:hard_elas}
\end{figure}

At the interface between the phase space of large-angle scattering ($\mathcal{C}_{\textrm{large-angle}}^{2\leftrightarrow2}$) and splitting-like processes ($\mathcal{C}_{\textrm{split}}^{2\leftrightarrow2}$), the two collision kernels should be consistent.
We verified this in Fig.~\ref{fig:hard_elas}: the differential rates of $\mathcal{C}_{\textrm{split}}^{2\leftrightarrow2}$ and $\mathcal{C}_{\textrm{large-angle}}^{2\leftrightarrow2}$ are compatible in $\sqrt{3\pcut T} < \omega < \pcut$. As long as we choose the cutoff $\Lambda$ in this range, this division of $\mathcal{C}^{2\leftrightarrow 2}_{\textrm{hard}}$ should be consistent.

A detailed discussion of the splitting approximation process is in \app{app:splittingapprox}. 
The  $p \gg T$ and $\omega \gg T$ kinematic cuts lead to significant simplifications for the matrix elements entering into $\mathcal{C}_{\textrm{split}}^{2\leftrightarrow2}$.

\subsubsection{Treatment of soft interactions}

\label{sec:hard_soft_theory:reformulation:soft}

In the hard-soft factorized approach, the large number of  soft interactions are described stochastically with drag and diffusion coefficients. When the momentum transfer is small, the Boltzmann equation [Eq. (\ref{eq:Boltzmann})] can be approximated as a Fokker-Planck equation. The collision kernel of the Fokker-Planck equation is written as: 
\begin{eqnarray}
\mathcal{C}_{\textrm{diff}}^{1\leftrightarrow 2, 2\leftrightarrow2} & = & \mathcal{C}^{1\leftrightarrow 2}_{\textrm{soft}}[\delta f]+\mathcal{C}^{2\leftrightarrow 2}_{\textrm{soft}}[\delta f] \nn \\
&= & -\frac{\partial}{\partial p^i}\left[\eta_{D, \textrm{soft}}p^i\delta f\right]-\frac{1}{2}\frac{\partial^2}{\partial p^i\partial p^j}\times \nn \\
& &\left\{\left[\hat{p}^i\hat{p}^j \hat{q}_{L, \textrm{soft}}+\frac{1}{2}\left(\delta^{ij}-\hat{p}^i\hat{p}^j\right)\hat{q}_\textrm{soft}\right]\delta f\right\} \nn \, , \\
    \label{eq:FP}
\end{eqnarray}
where $\eta_{D, \textrm{soft}}$ is the drag coefficient of the soft interactions, $\hat{q}_{L, \textrm{soft}}$ and $\hat{q}_\textrm{soft}$ are the longitudinal and transverse momentum diffusion coefficients of the soft interactions. 

In the diffusion process, the number and the identity of the particles are preserved. Since the soft radiations of the $1 \leftrightarrow 2$ process are absorbed by the plasma, the number of particles is also preserved.
We include both the soft $1\leftrightarrow 2$ and $2\leftrightarrow 2$ collisions in the diffusion process. The diffusion process can be solved using a Langevin equation~\cite{He:2013zua} in the numerical implementation. %

For soft $1\leftrightarrow 2$ process, we can obtain the perturbative longitudinal diffusion coefficients by expanding $\mathcal{C}^{1\leftrightarrow 2}$ and only keeping the soft radiation terms. At leading order in $\alpha_s$,
\begin{equation}
\hat{q}^{1\leftrightarrow 2}_{L, \textrm{soft}} = \frac{(2-\ln2)g^4C_RC_AT^2\mu_\omega}{4\pi^3} \, ,
\label{eq:inel_qhat_L}
\end{equation}
where $C_R$ is the Casimir factor. For gluons, $C_R=C_A$, while for quarks, $C_R=C_F$.\footnote{\label{footnote:gluons_low_omega} Note that the diffusion coefficient $\hat{q}^{1\leftrightarrow 2}_{L, \textrm{soft}}$ does not depend on the number of the quark flavor, because very soft radiations are dominated by gluon scatterings.} The derivation of this value can be found in Appendix~\ref{appendix:inel_low_omega}. 

We assume that the radiation angle is zero for collinear radiations. Consequently the transverse diffusion coefficient of $1\leftrightarrow 2$ interactions is approximated as zero.

For soft $2\leftrightarrow 2$ processes, 
the diffusion coefficients can be calculated perturbatively;
a modern derivation  
can be found in Ref.~\cite{Ghiglieri:2015ala}.
The transverse momentum diffusion coefficient due to soft scatterings is
\begin{equation}
   \hat{q}_{\rm soft}^{2 \leftrightarrow 2} 
   = \frac{g^2 C_R T m_D^2}{4\pi} \ln \left[1+\left(\frac{\mu_{\tilde{q}_\perp}} {m_D}\right)^2\right] \, ,
    \label{eq:elas_qhat}
 \end{equation}
where 
$m_D^2 \equiv g^2T^2(N_c/3+N_f/6)$ is the square of the leading order Debye mass, $N_c=3$ is the number of colors and $N_f$ is the number of flavors involved in the interactions. 
The longitudinal diffusion coefficient at order $\mathcal{O}(\alpha_s)$ is
\begin{equation}
    \hat{q}^{2\leftrightarrow 2}_{L, \textrm{soft}} = \frac{g^2C_RTM_\infty^2}{4\pi}\ln\left[1+\left(\frac{\mu_{\tilde{q}_\perp}}{M_\infty}\right)^2\right]\, ,
    \label{eq:elas_qhat_L}
\end{equation}
where $M_\infty \equiv \sqrt{m_D^2/2}$ is the gluon asymptotic thermal mass~\cite{Blaizot:2001nr,Ghiglieri:2015ala}.

Since detailed balance is preserved in the Fokker-Planck equation, as verified in  \app{app:detailed_balance}, the drag coefficient $\eta_D$ can be calculated from diffusion coefficients according to Einstein relation for both soft $1\leftrightarrow 2$ and $2\leftrightarrow 2$ processes: 
\begin{equation}
\eta_{D, \textrm{soft}}(E) = \frac{\hat{q}_{L, \textrm{soft}}}{2Tp}\left[1+\mathcal{O}\left(\frac{T}{p}\right)\right]\, . 
\label{eq:drag}
\end{equation}

Equations (\ref{eq:inel_qhat_L}-\ref{eq:drag}) assume that the coupling $\alpha_s$ is small. We discuss the range of validity of the perturbative coefficients in Section \ref{sec:weak_coupling_analytical_vs_numerical_qhat}. 
Our long-term goal is to treat 
$\hat{q}_{\rm soft}$ 
and 
$\hat q_{L,\rm soft}$
as non-perturbative parameters, incorporating
much more physics than leading order %
scattering. These parameters could then either be constrained with lattice inputs~\cite{Moore:2019lgw} or fitted to experimental data, e.g. with the Bayesian approach~\cite{JETSCAPE:2021ehl,Ke:2020clc}. In either case, the results will depend on the separation scale
$\mu$, 
and this dependence 
would then have to match with the hard sector at LO (order $g^2$), NLO (order $g^3$), and NNLO (order $g^4$, the first order the coupling runs).
Ideally the hard sector, and thus the evolution with
$\mu$ 
can be treated perturbatively. 
As a first step we will study the sensitivity to the scale separation 
$\mu$
in this manuscript.
Besides the identity preserving diffusion process, the identity of the particle can be converted through soft fermion
exchange with the medium. This exchange must be screened with the non-perturbative HTL resummation scheme. In the hard-soft factorized approach adopted here, we separate the $2\leftrightarrow 2$ processes with fermion exchange into 
hard  collisions with $\tilde q_\perp > \mu_{\tilde q_\perp}$,  and soft collisions with $\tilde q_\perp < \mu_{\tilde q_\perp}$ (see \Fig{fig:reformulation}). The hard exchange collisions are treated with vacuum matrix elements, while the soft exchange collisions are incorporated into  
a conversion rate $\Gamma_{q \rightarrow g}^{\rm conv}(p)$ for $q \rightarrow g$: %
\begin{equation}
\Gamma_{q\rightarrow g}^{\rm conv}(p) =  \frac{g^2C_F m_\infty^2 }{16\pi p} \log\left[1 + \frac{\mu_{\tilde q_\perp}^2}{m_{\infty}^2} \right] \, .
\end{equation}
Here $m_{\infty}^2$ is the fermion asymptotic mass, $m_{\infty}^2 = g^2 C_F T^2/4$~\cite{Blaizot:2001nr,Ghiglieri:2015ala}.
In each time step there is a probability $\Delta t \, \Gamma^{\rm conv}$ for a quark to become a gluon, with the same momentum, and vice versa.
Further details about the conversion rate $C_{\rm conv}^{2\leftrightarrow 2}$ are given in \app{app:soft_conv}.
In the future, the non-perturbative conversion coefficient $\Gamma_{q \rightarrow g}^{\rm conv}$ can be taken from 
a next-to-leading order analysis~\cite{Ghiglieri:2015ala}, 
or can be determined from data in a Bayesian approach.
\subsubsection{Summary}

In summary, the collision kernel of hard-soft factorized model is reformulated as
\begin{eqnarray}
\mathcal{C} &=& \mathcal{C}^{2\leftrightarrow 2} + \mathcal{C}^{1\leftrightarrow 2} \nn \\
&=& \mathcal{C}_{\textrm{large-}\omega}^{1\leftrightarrow2}\left(\mu_\omega\right)+\mathcal{C}_{\textrm{large-angle}}^{2\leftrightarrow2}\left(\mu_{\tilde{q}_\perp}, \Lambda\right)+ \mathcal{C}_{\textrm{split}}^{2\leftrightarrow2}\left(\Lambda\right) \nn \\
& &\;\;\;\; +\mathcal{C}_{\textrm{diff}}^{1\leftrightarrow2, 2\leftrightarrow2}\left(\mu_\omega, \mu_{\tilde{q}_\perp}\right)+\mathcal{C}^{2\leftrightarrow 2}_{\textrm{conv}}\left(\mu_{\tilde{q}_\perp}\right) \, .
\label{reformulation}
\end{eqnarray}

The cutoff dependence of the stochastic description is cancelled in Eq.~(\ref{reformulation}) by the cutoff dependence of the hard interactions.
That is, each individual process in the hard-soft factorized model is dependent on the cutoff, but this dependence cancels out when all the processes are summed. 
We show this explicitly in Section~\ref{sec:tequila_weak_coupling}.

\subsection{Running of the strong coupling $\alpha_s$}

All discussions up to this point assumed that the strong coupling constant $\alpha_s$ is fixed at a given small value. 
It is clear, however, that the strong coupling constant will be different for soft and hard interactions; this is in fact a key assumption of the present model: hard interactions are more perturbative than soft ones, because the coupling constant scales inversely with the momentum exchange between the energetic parton and the plasma (see Ref.~\cite[Section V]{CaronHuot:2010bp} for a discussion, for example).
The running is slow (logarithmic in the momentum exchange), however, more studies will be necessary to understand the exact magnitude of loop corrections or non-perturbative effects on soft and hard collisions.

As a first step in introducing our model of parton energy loss, we keep the strong coupling constant $\alpha_s$ fixed throughout the manuscript.

\section{Hard-soft factorization of parton energy loss in the weakly-coupled regime: numerical study}

\label{sec:tequila_weak_coupling}

In the first part of this section, we compare the analytical equations for the soft-interaction parton transport coefficients [Eqs.~(\ref{eq:inel_qhat_L}--\ref{eq:elas_qhat_L})] with their numerical values evaluated from the matrix elements, and summarize the range of cutoff and coupling where they are consistent. We also compare (i) soft interactions modelled with matrix elements with (ii) soft interactions modelled with the Langevin equation. We perform this test in the weak coupling limit. We use this discussion to review the range of validity of the Fokker-Planck equation and its stochastic Langevin implementation. 

In the second part of this section, we compute the energy loss of an energetic parton in a brick and discuss the dependence of the results on the soft-hard cutoffs introduced in Section~\ref{sec:factorized_model}.

\subsection{Treatment of soft interactions}
\label{sec:treatment_of_soft_interactions}
Soft interactions can be described either stochastically with transport coefficients, or microscopically with matrix elements.
In what follows, we compare these two descriptions,
with particular emphasis on the effect of the soft-hard cutoffs and of the coupling constant.

The tests performed in the present subsection are not expected to be related to exact composition of the plasma (number of quark flavors). Thus, for simplicity, the calculations are performed in the pure glue limit ($N_f=0$).

\subsubsection{Analytical and numerical calculation of soft transport coefficients}
\label{sec:weak_coupling_analytical_vs_numerical_qhat}

In a weakly-coupled quark-gluon plasma, the  drag and diffusion coefficients for soft interactions can be calculated analytically using perturbation theory [~Eqs.(\ref{eq:inel_qhat_L}-\ref{eq:elas_qhat_L})~], as discussed in Section~\ref{sec:hard_soft_theory:reformulation:soft}.
The same drag and diffusion coefficients can be obtained by direct numerical integration of the parton energy loss rates; these rates are calculated from matrix elements screened by plasma effects~\cite{Arnold:2003zc}.

The diffusion coefficients are defined as \cite{Ghiglieri:2015ala}
\begin{equation}
\begin{split}
    \hat{q}(p) &\equiv \frac{d}{dt}\left<\left(\Delta p_\perp\right)^2\right>, \\
    \hat{q}_L(p) &\equiv \frac{d}{dt}\left<\left(\Delta p_L\right)^2\right>,
\end{split}
    \label{eq:df_qhat}
\end{equation}
where $\Delta p_\perp$ is the momentum change perpendicular to the direction of the energetic parton, and $\Delta p_L$ is the longitudinal momentum change of the parton. The brackets represent an average over all interactions during the parton propagation. The numerical soft diffusion rates are thus calculated as \cite{Ghiglieri:2015ala}
\begin{equation}
\begin{split}
    \hat{q}^{2\leftrightarrow 2}_{\textrm{soft}}(p) &= \int_0^{\mu_{\tilde{q}_\perp}} d\tilde{q}_\perp \int_{-\infty}^{\Lambda} d\omega \tilde{q}_\perp^2 \left.\frac{d^2\Gamma(\bm{p}, \bm{q})}{d\omega d\tilde{q}_\perp}\right|_{2 \leftrightarrow 2}, \\
    \hat{q}^{2\leftrightarrow 2}_{L, \textrm{soft}}(p) &= \int_0^{\mu_{\tilde{q}_\perp}}d\tilde{q}_\perp \int_{-\infty}^{\Lambda} d\omega \omega^2\left.\frac{d^2\Gamma(\bm{p}, \bm{q})}{d\omega d\tilde{q}_\perp}\right|_{2 \leftrightarrow 2}, \\
    \hat{q}^{1\leftrightarrow 2}_{L, \textrm{soft}}(p) &= \int_{-\mu_\omega}^{\mu_\omega} d\omega \omega^2\left.\frac{d\Gamma(\bm{p}, \bm{q})}{d\omega}\right|_{1 \leftrightarrow 2},
\end{split}
\label{eq:num_qhat}
\end{equation}
where $d\Gamma(\bm{p}, \bm{q})/d\omega$ and $d^2\Gamma(\bm{p}, \bm{q})/d\omega d\tilde{q}_\perp$ are the rates for an energetic parton with four-momentum $(p,\bm{p})$ to undergo a four-momentum change $(\omega,\bm{q})$ calculated using screened matrix elements. The initial parton energy $p$ is assumed to be much larger than all other energy scales in the problem, effectively $p\to\infty$.
The cutoffs $\mu_{\tilde{q}_\perp}$, $\Lambda$ and $\mu_\omega$ are used to limit the phase space of interactions included in the transport coefficients, in the present case to limit the interactions to soft ones only.

\begin{figure}
    \centering
    \includegraphics[width=\linewidth]{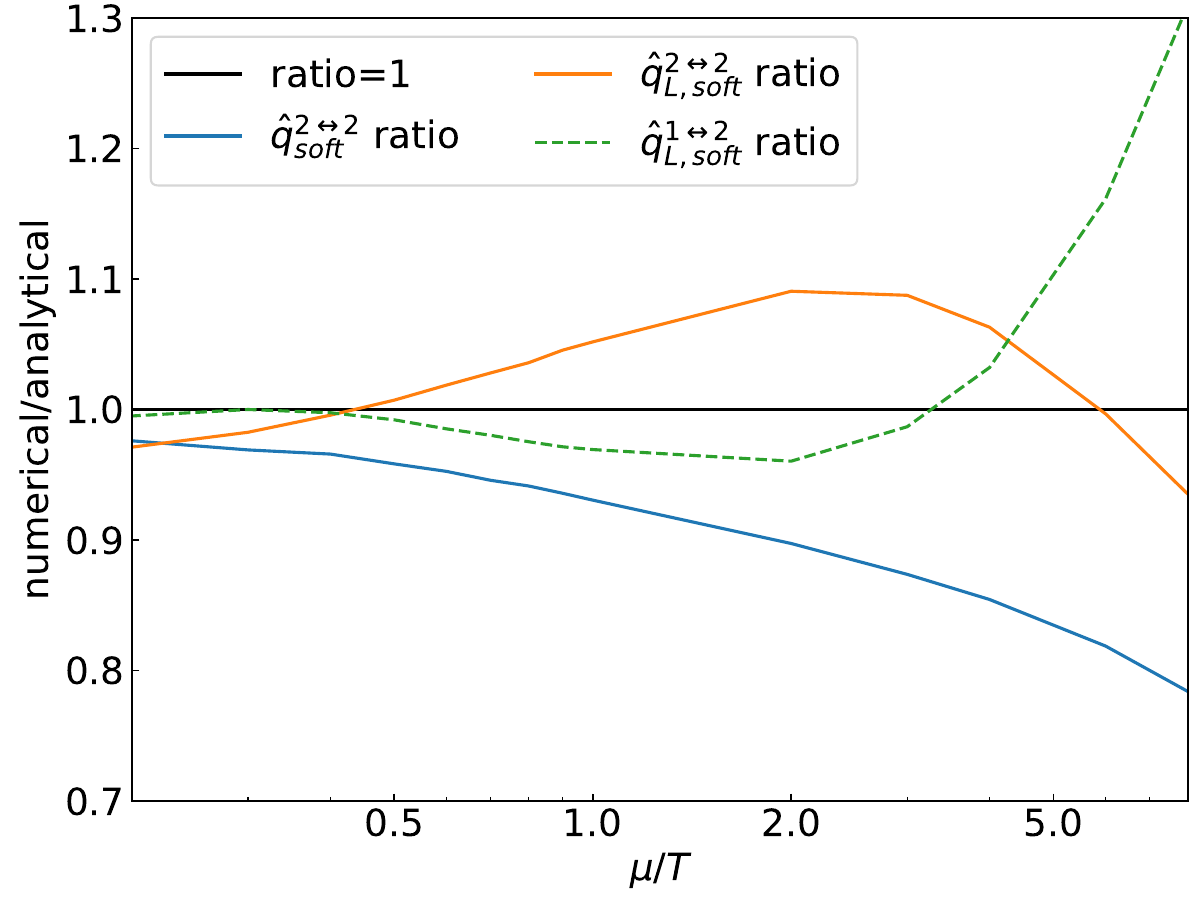}
    \caption{The ratio between the numerical (\eqref{eq:num_qhat}) and analytical (Eqs.(\ref{eq:inel_qhat_L}-\ref{eq:elas_qhat_L})) 
    momentum transport coefficients: $\hat{q}^{1\leftrightarrow 2}_{L, \textrm{soft}}$, $\hat{q}^{2\leftrightarrow 2}_{\textrm{soft}}$ and $\hat{q}^{2\leftrightarrow 2}_{L, \textrm{soft}}$. The numerical
    results are computed 
    with exact $1\leftrightarrow 2$ or $2\leftrightarrow 2$ kinematics up to a cutoff $\mu/T$. The analytical coefficients make kinematic approximations appropriate for $\mu/T \ll 1$.
    The results are shown for different values of the hard-soft cutoffs at $\alpha_s=0.005$. We calculate these results using $p_0 = 100$~GeV and $T=300$~MeV in a pure glue medium ($N_f=0$). The cutoff $\mu$ in the figure denotes $\mu_{\tilde{q}_\perp}$ for $\hat{q}^{2\leftrightarrow 2}_{\textrm{soft}}$ and $\hat{q}^{2\leftrightarrow 2}_{L, \textrm{soft}}$, and $\mu_\omega$ for $\hat{q}^{1\leftrightarrow 2}_{L, \textrm{soft}}$. In the elastic case, the additional cutoff on $\omega$ is set to $\Lambda=\min(p_{\textrm{cut}}, \sqrt{3p_0T})$. }
    \label{fig:qhat_cutoff}
\end{figure}

There are two important differences between Eq.~(\ref{eq:num_qhat}) and the analytical diffusion coefficients Eqs.~(\ref{eq:inel_qhat_L}-\ref{eq:elas_qhat_L}). 
First, Eq.~(\ref{eq:num_qhat}) is formally valid for arbitrarily large cutoffs ($\mu_{\tilde{q}_\perp}$, $\Lambda$ and $\mu_{\omega}$), while Eqs.~(\ref{eq:inel_qhat_L}- \ref{eq:elas_qhat_L}) assume the cutoff to be at most of order $T$.
Second, there is the question of the smallness of the coupling. Equations~(\ref{eq:inel_qhat_L}- \ref{eq:elas_qhat_L}) are derived assuming $\alpha_s\ll 1$. Equation~(\ref{eq:num_qhat}) is valid at arbitrarily coupling, although the rates $d\Gamma(\bm{p}, \bm{q})/d\omega$ and $d^2\Gamma(\bm{p}, \bm{q})/d\omega d\tilde{q}_\perp$ themselves are typically calculated perturbatively.\footnote{It is highlighted in Ref.~\cite{CaronHuot:2010bp} that the AMY differential equation used to evaluate the inelastic collisions rate remains similar if interactions with the plasma are non-perturbative. One difference is the perturbative partonic collision kernel $C(\mathbf{q}) \propto m_D^2 /\left[ \mathbf{q}^2 (\mathbf{q}^2+m_D^2) \right]$ that must be modified. Non-perturbative contributions to the thermal masses are another difference.\label{footnote:caronhuot}}

A comparison of Eq.~(\ref{eq:num_qhat}) and the analytical diffusion coefficients Eqs.~(\ref{eq:inel_qhat_L}-\ref{eq:elas_qhat_L}) is shown in Fig.~\ref{fig:qhat_cutoff} as a function of the different cutoffs. This comparison is made at weak coupling ($\alpha_s=0.005$) and yields the expected agreement between the two approaches, as long as the cutoffs are $\lesssim T$.

\begin{figure}
    \centering
    \includegraphics[width=\linewidth]{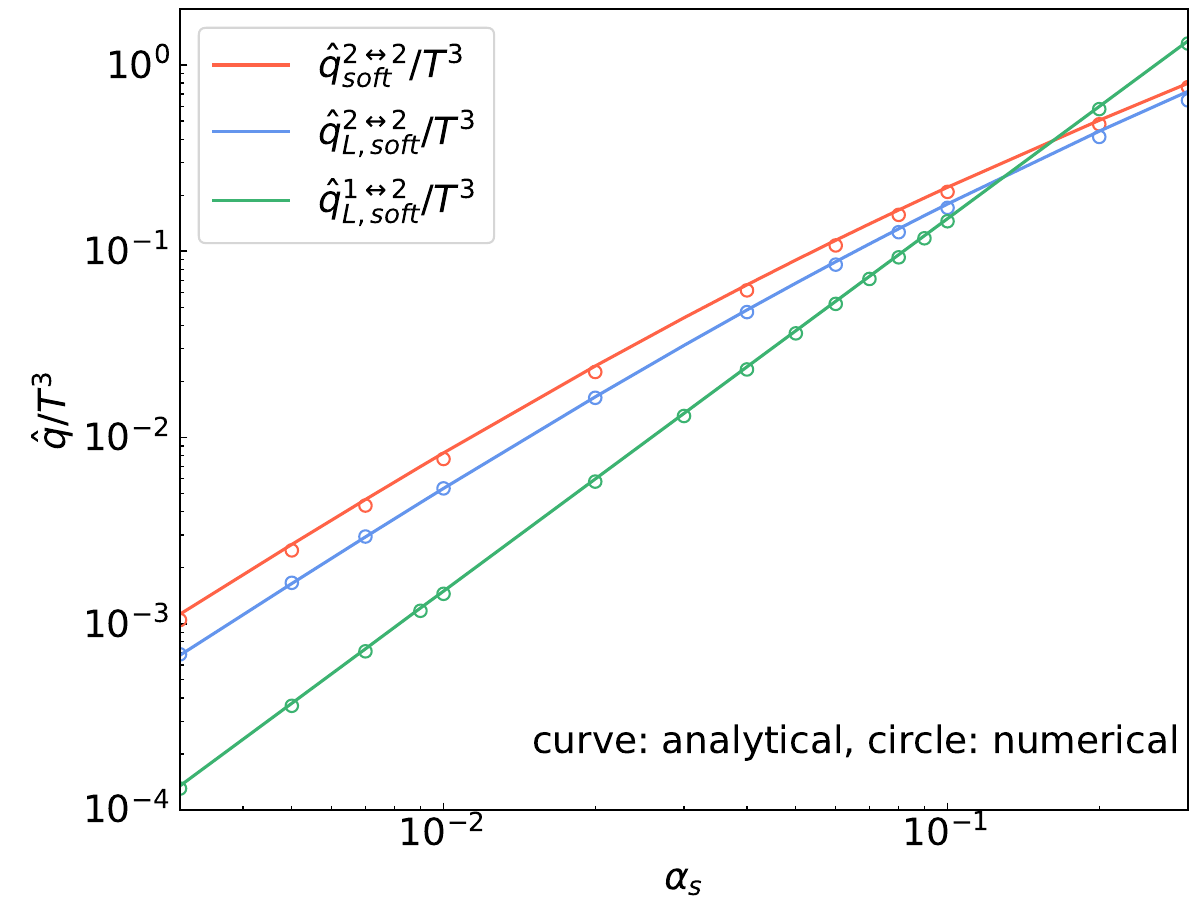}
    \caption{Comparison of the numerical and analytical $\hat{q}^{1\leftrightarrow 2}_{L, \textrm{soft}}$, $\hat{q}^{2\leftrightarrow 2}_{\textrm{soft}}$ and $\hat{q}^{2\leftrightarrow 2}_{L, \textrm{soft}}$ with different coupling constants $\alpha_s$ (see Fig.~\ref{fig:qhat_cutoff} for description). The solid curves denote analytical results, and the circles denote numerical results. For the kinematic cutoffs, we use $\mu_{\tilde{q}_\perp} = \mu_\omega = T$ and $\Lambda=\min(\sqrt{3p_0T}, \pcut)$. The numerical values of the transport coefficients were calculated assuming a $T=300$~MeV pure glue medium ($N_f=0$) and an energetic parton with $p_0 = 100$~GeV. }
    \label{fig:qhat_coupling}
\end{figure}

In Fig.~\ref{fig:qhat_coupling} we compare the analytical soft diffusion coefficients Eqs.~(\ref{eq:inel_qhat_L}-\ref{eq:elas_qhat_L}) with the numerical soft diffusion coefficients Eq.~(\ref{eq:num_qhat}) at different values of the strong coupling constant $\alpha_s$. 
We find that the analytical soft diffusion coefficients agree well with the numerical calculations even at large coupling, except for a small tension in $\hat{q}^{2\leftrightarrow 2}_{L, \textrm{soft}}$ at large $\alpha_s$.
Tension between different calculations of the soft transport coefficients are in fact not unexpected: perturbative calculations can be equivalent at order $g^n$ yet be different at order $g^{n+1}$. These differences are negligible at weak coupling, but can become significant for larger values of the coupling. This is a natural consequence of pushing the calculations beyond their regimes of validity. 
There is a practical consequence: two parton energy loss calculations that use the exact same approach (weakly-coupled kinetic theory) can lead to different results, when used at large coupling; neither approach is more ``correct'' than the other. This is important to keep in mind when comparing the present soft-hard factorized energy loss model with other implementations such as Ref.~\cite{Schenke:2009gb}.

\subsubsection{Theoretical guidance on the range of applicability of the Fokker-Planck equation}

\label{sec:hard_soft_at_weak_coupling:soft_interactions:scale}

The energy loss of energetic partons through soft interactions is described by solving the Fokker-Planck equation with a stochastic Langevin approach. The applicability of the stochastic description is limited to the regime where the Fokker-Planck equation holds. 
This regime of applicability depends partly on properties of the interactions rates.
We can summarize the regime of validity of the Fokker-Planck equation by first expanding the \emph{Boltzmann equation} for soft collisions (around $\omega=0$): 
\begin{equation}
\begin{split}
    \partial_t f(p, t) = &\left<\omega\right> f^{(1, 0)}(p, t) +\frac{1}{2}\left<\omega^2\right>f^{(2, 0)}(p, t)\\ 
    &+\frac{1}{6}\left<\omega^3\right>f^{(3, 0)}(p, t)+\dots, 
\end{split}
\label{eq:boltzmann_small_w_expansion}
\end{equation}
where $f(p,t)$ is the momentum distribution of energetic partons at time $t$ and
\begin{equation}
    \left<\omega^k\right> =  \int d\omega \omega^k\frac{d\Gamma}{d\omega}
    \label{eq:rate_moment}
\end{equation}
is the k-th moment of the differential collision rate $d\Gamma/d\omega$.\footnote{
The bounds on the integration are the same as in Eqs.~(\ref{eq:num_qhat}), including the additional integration over $\tilde{q}_\perp$ necessary in the elastic case.}

By keeping only the first two terms on the right-hand side, \eqref{eq:boltzmann_small_w_expansion} simplifies to the Fokker-Planck equation. 

Assuming a single initial energetic parton of energy $p_0$,
\begin{equation}
    f(p,t=0)=\delta(p-p_0),
\end{equation}
the solution of the Fokker-Planck equation is
\begin{equation}
    \begin{split}
        f_{FP}(p, t)&=\frac{\exp{\left[-\frac{\left( p-(p_0-\left<\omega\right>t ) \right)^2 }{2t\left<\omega^2\right>}\right]}}{\sqrt{2\pi\left<\omega^2\right> t}} \, .
    \end{split}
    \label{eq:toy_FP}
\end{equation}
The above solution simply describes the energy distribution of the energetic parton widening from scatterings with $\hat{q}_L = \left<\omega^2\right>$ energy diffusion, and an average energy loss of $\left<\omega\right> t$.

Using this solution, we can compute the ratio of the third and second terms in the expanded Boltzmann equation (\eqref{eq:boltzmann_small_w_expansion}): 
\begin{equation}
    \mathcal{R} = \frac{\frac{1}{6}\left<\omega^3\right>f_{FP}^{(3, 0)}(p, t)}{\frac{1}{2}\left<\omega^2\right>f_{FP}^{(2, 0)}(p, t)} = -\frac{2\Delta p \left<\omega^3\right>(\Delta p^2-3\left<\omega^2\right>t)}{3\left<\omega^2\right>t(\Delta p^2-\left<\omega^2\right>t)} \, ,
    \label{eq:R_factor}
\end{equation}
where $\Delta p = p-(p_0-\left<\omega\right>t)$ is the distance in momentum from the peak of the Fokker-Planck solution (\eqref{eq:toy_FP}). 

Significant corrections to the Fokker-Planck solution \eqref{eq:toy_FP} are expected unless $\mathcal{R} \ll 1$.
As is clear from \eqref{eq:R_factor}, the range of validity of the Fokker-Planck equation depends on properties of the rate (the second and third moments $\left<\omega^2\right>$ and $\left<\omega^3\right>$), as well as on time $t$ and on the distance in momentum $\Delta p$ from the peak of the distribution.

The Fokker-Planck equation describes the effect of soft interactions on an energetic parton. 
The soft interactions dominate for small values of $\Delta p$.
Expanding $\mathcal{R}$ (\eqref{eq:R_factor}) around $\Delta p=0$, we obtain:
\begin{equation}
    \mathcal{R} = -\frac{\Delta p \left<\omega^3\right>}{\left<\omega^2\right>^2 t}+\frac{2\Delta p^3\left<\omega^3\right>}{3\left<\omega^2\right>^3t^2}+\mathcal{O}(\Delta p^5) \, .
    \label{eq:R_factor_expansion}
\end{equation}

By taking the ratio of the second and first term of this expansion, 
\begin{equation}
r \equiv \frac{-2\Delta p^3\left<\omega^3\right>}{3\left<\omega^2\right>^3t^2} \bigg/ \frac{\Delta p \left<\omega^3\right>}{\left<\omega^2\right>^2 t},
\end{equation}
we can find the value of $\Delta p$ for which this ratio will be large:
\begin{equation}
\Delta p = \sqrt{\frac{3}{2}r}\sqrt{\left<\omega^2\right>t} \, ,
\label{eq:Delta_p}
\end{equation}
with $r$ a constant assumed to be smaller than $1$.
We can use this value of $\Delta p$ as the range of momentum around the mean energy loss that can reasonably be described by the Fokker-Planck equation. 
Using \eqref{eq:Delta_p} and the first term of \eqref{eq:R_factor_expansion}, we define the scale $\mathcal{S}$ as
\begin{equation}
    \mathcal{S}=\frac{\left<\omega^3\right>}{\left<\omega^2\right>^{3/2}} \frac{1}{\sqrt{t}} \, .
    \label{eq:scale}
\end{equation}

When this scale $\mathcal{S}$ is much smaller than 1, the Fokker-Planck equation is expected to provide a good description of the Boltzmann equation in the relevant range of momentum. We emphasize that Eq.~(\ref{eq:scale}) was derived without any specific form for the rate $d\Gamma/d\omega$; in particular, the formula is the same for perturbative and non-perturbative calculations of the rate.

\paragraph{Scale for inelastic rate}

For inelastic interactions at weak coupling, we can evaluate Eq.~\ref{eq:scale} analytically using the formula for the very soft inelastic differential rate described in Eq. (\ref{eq:inel_soft_rate}). In this soft inelastic limit, the scale is given by
\begin{equation}
\mathcal{S}_{1\leftrightarrow 2}=\frac{\pi ^{3/2}}{3 C_A \sqrt{2-\ln(2)}} \frac{\mu_\omega^{3/2}}{g^2  T^2 \sqrt{t}}.
\label{eq:scale_inel}
\end{equation}
This implies that soft inelastic emissions with energy smaller than $\mu$ can be described with the Langevin equation as long as the evolution time $t$ in the medium is sufficiently long:
\begin{equation}
    t \gg \frac{\mu_\omega^3}{g^4 T^4}.
    \label{eq:time_langevin_inel}
\end{equation}
Assuming $\mu_\omega \sim T$ results in $t \gg 1/[g^4 T]$, while $\mu_\omega \sim g T$ results in $t \gg 1/[g T]$. 
This implies that there is a very large difference between a stochastic description of soft interactions with $\omega \lesssim T$ compared to soft interactions with $\omega \lesssim g T$: in the former case, one needs a plasma $1/g^3$ larger.
These values serve as a reminder that, while one can in principle increase the phase space of interactions described stochastically,
one may need an unrealistically large plasma for this description to be valid.

\paragraph{Scale for elastic rate}

\label{sec:elastic_scale}

\begin{figure}
    \centering
    \includegraphics[width=\linewidth]{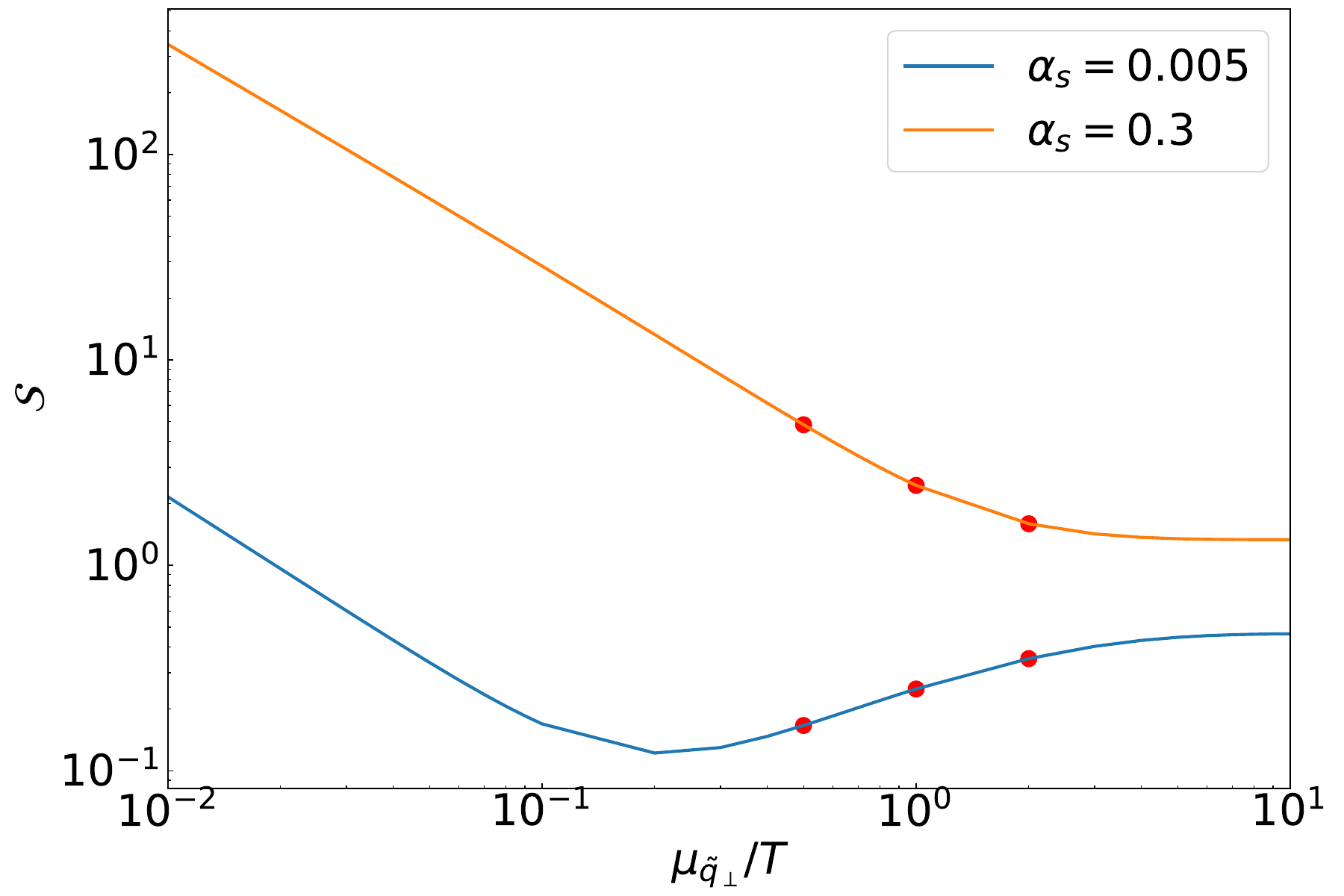}
    \caption{Dependence of the skewness scale $\mathcal{S}$ on the cutoff $\mu_{\tilde{q}_\perp}$, for the elastic parton energy loss rate. The top line is for $\alpha_s=0.3$ and the bottom line for $\alpha_s=0.005$. The points denote the values corresponding to $\mu_{\tilde{q}_\perp} = 0.5, 1, 2T$. This interaction rate is calculated assuming a pure glue medium ($N_f = 0$). }
    \label{fig:elastic_scale_num}
\end{figure}

The dependence of the scale $\mathcal{S}$ [Eq.~(\ref{eq:scale})] on the cutoff $\mu_{\tilde{q}_\perp}$ is shown in  Fig.~\ref{fig:elastic_scale_num}, for a small and large value of the coupling constant: $\alpha_s=0.005 \textrm{ and } 0.3$. One can see that the dependence on the cutoff can be non-monotonic for small values of $\alpha_s$, unlike in the inelastic case.
Numerical tests, as well as the analytical expression available for the second moment at small coupling [Eq.~(\ref{eq:elas_qhat_L})], suggest that the second moment of the elastic rate is the origin of this non-monotonic dependence of the elastic scale $\mathcal{S}$ on $\mu_{\tilde{q}_\perp}$.

\subsubsection{Comparison between the diffusion process and the collision rate}

\label{sec:weak_coupling_diffusion_vs_rate}

\begin{figure*}
    \centering
    \includegraphics[width=2\columnwidth]{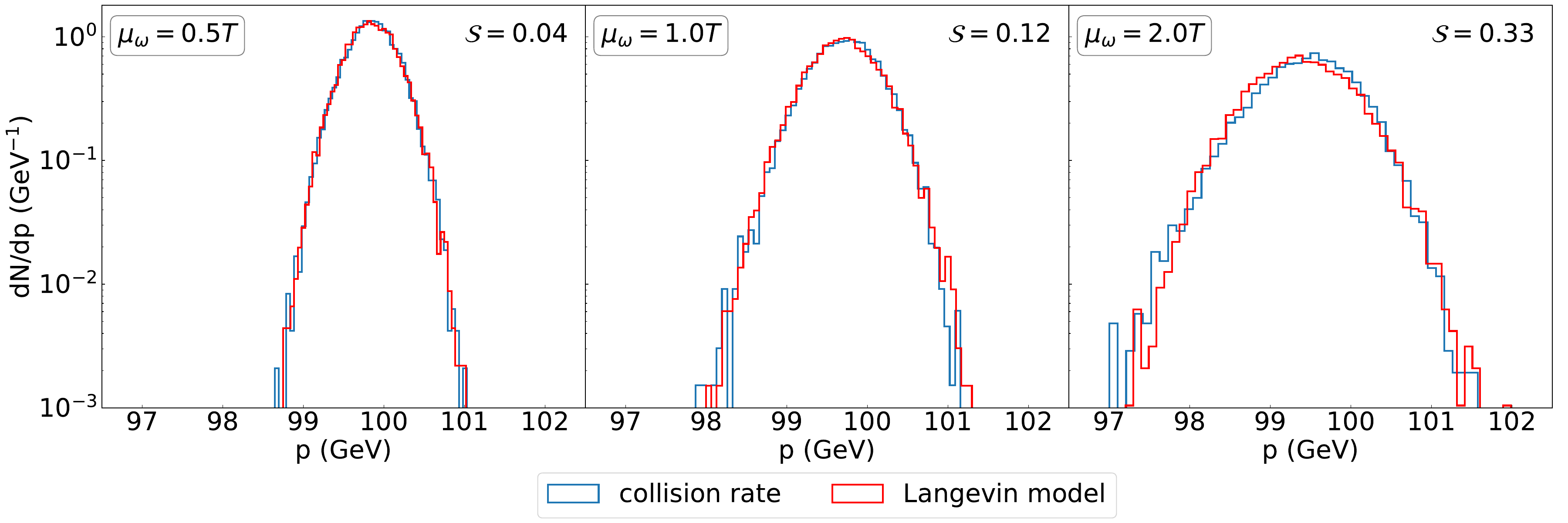}
    \caption{The energy distribution of a $100$~GeV gluon propagating through a $300$~MeV pure glue medium ($N_f = 0$) at $\alpha_s = 0.005$. The evolution time is $t = (0.3/\alpha_s )^2=3600$~fm. Only soft $1\leftrightarrow 2$ interactions with $\omega < \mu_\omega$ are allowed. Three different values of the cutoff are shown: $\mu_\omega/T=0.5, 1, 2$.}
    \label{fig:diffusion_collision_rate_small_coupling:inel}
\end{figure*}

\begin{figure*}
    \includegraphics[width=2\columnwidth]{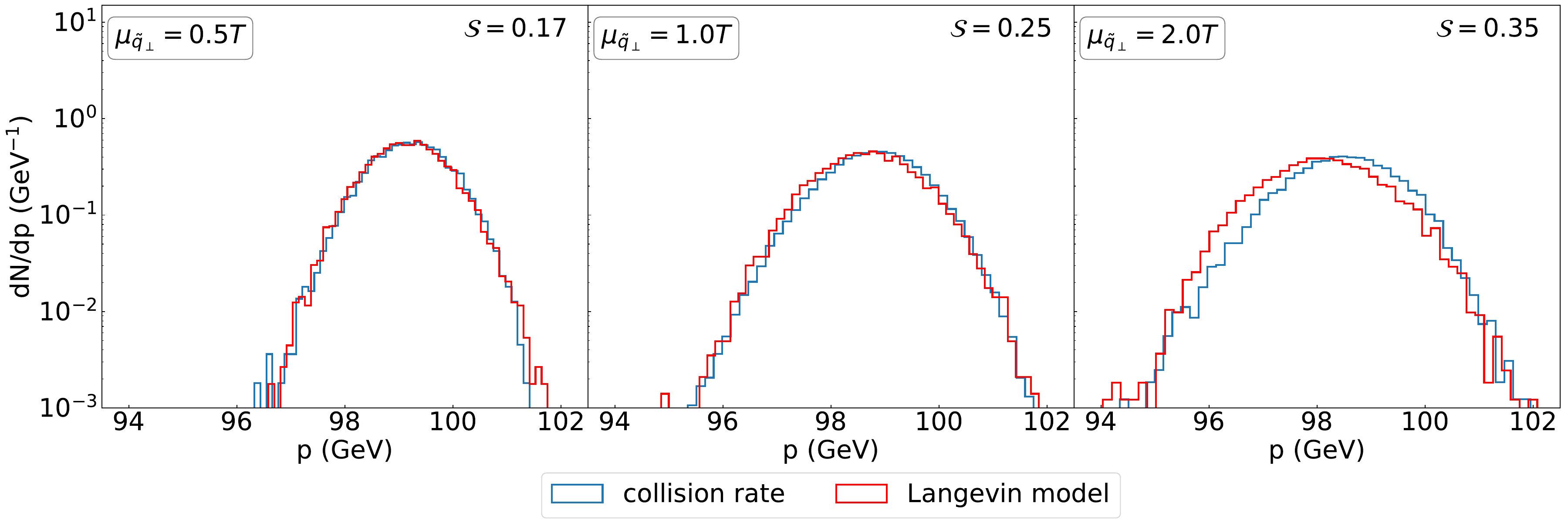}
    \onecolumngrid
    \caption{The energy distribution of a $100$~GeV gluon propagating through a $300$~MeV pure glue medium ($N_f = 0$) at $\alpha_s = 0.005$. The evolution time is $t = (0.3/\alpha_s )^2=3600$~fm. Only soft $2\leftrightarrow 2$ interactions with $\omega < \Lambda$ and $\tilde{q}_\perp < \mu_{\tilde{q}_\perp}$ are allowed. Three different values of the cutoff $\mu_{\tilde{q}_\perp}$ are shown: $\mu_{\tilde{q}_\perp}/T=0.5, 1, 2$. We choose $\Lambda=\min(p_{\textrm{cut}}, \sqrt{3p_0T})$. }
    \label{fig:diffusion_collision_rate_small_coupling:elastic}
\end{figure*}

In this section, we verify numerically the conclusion from the previous section:
we compare a stochastic and a microscopic evolution of energetic partons in a static medium. In the microscopic rate-based picture, we use kinematic cuts to forbid hard interactions of the energetic parton. 
Because we are comparing soft interactions, we must use screened elastic matrix elements~\cite{Arnold:2003zc} in the microscopic description.\footnote{Note that this is for testing purpose only, and that this is different from the vacuum matrix elements used for $\mathcal{C}^{2\leftrightarrow 2}_{\textrm{hard}}$ in the hard-soft factorized energy loss model.} 
The screened inelastic ($1\leftrightarrow 2$) rate is obtained numerically by solving the AMY differential equation, except for very small $\omega$ values, in which case the analytical expression described in Appendix~\ref{appendix:inel_low_omega} (\eqref{eq:inel_soft_rate}) is used.

We choose the hard-soft cutoffs (i.e. $\mu_\omega$ and $\mu_{\tilde{q}_\perp}$) to be at the order of $T$ in the following tests.
We set the coupling to be $\alpha_s = 0.005$, which corresponds to $g\approx 0.25$.
We choose $T=300$~MeV for the temperature of the fluid, and set the propagation time in the plasma to be $t=(0.3/\alpha_s)^2=3600$~fm.\footnote{We choose the evolution time $t \propto 1/\alpha_s^2$ to keep the number of the collisions approximately the same for different values of $\alpha_s$. With the choice $t=(0.3/\alpha_s)^2$, the evolution time is $1$~fm when we use $\alpha_s=0.3$ later in the manuscript.}

We perform the diffusion approach and the collision rate approach separately to calculate the single parton energy distribution of a hard $100$~GeV gluon propagating in the static pure glue medium. We emphasize once again that we only include soft interactions in the test by introducing the following hard-soft cutoffs on radiation energy and momentum transfer: for $\mathcal{C}^{1\leftrightarrow 2}_{\textrm{soft}}$, we only include radiations with the radiation energy $\omega < \mu_\omega$; while for $\mathcal{C}^{2\leftrightarrow 2}_{\textrm{soft}}$, we only include interactions with $\tilde{q}_\perp < \mu_{\tilde{q}_\perp}$ and the energy transfer $\omega < \Lambda$. 

According to \eqref{eq:scale}, for inelastic interactions ($\mathcal{C}^{1\leftrightarrow 2}_{\textrm{soft}}$) to be describable stochastically for a cutoff $\sim T$, one needs $t \gg 1/[g^4 T] \approx 200$~fm of propagation time in the conditions described above. As expected, we find in Figure \ref{fig:diffusion_collision_rate_small_coupling:inel} that for inelastic interactions, in the weakly-coupled regime, the diffusion process can reproduce the single parton energy distribution generated by the collision-rate process. The value of the scale $\mathcal{S}$, shown for each cutoff $\mu_\omega$, are indeed smaller than $1$. As $\mu_\omega$ increases, small differences appear between the Langevin description and the microscopic collision approach; the scale $\mathcal{S}$ is correspondingly larger, though still smaller than $1$.

The same results are shown for the elastic case ($\mathcal{C}^{2\leftrightarrow 2}_{\textrm{soft}}$) in Fig.~\ref{fig:diffusion_collision_rate_small_coupling:elastic}.  This time, the scale $\mathcal{S}$ is somewhat larger, and somewhat larger differences can indeed be seen between the Langevin and collision rate descriptions. As for the elastic case, the scale $\mathcal{S}$ increases as the cutoff increases, where more and more collisions are described stochastically.

\subsection{Parton energy loss at small coupling in a static medium}
\label{sec:parton_evolution_small}
\begin{figure*}
    \centering
     \includegraphics[width=2\columnwidth]{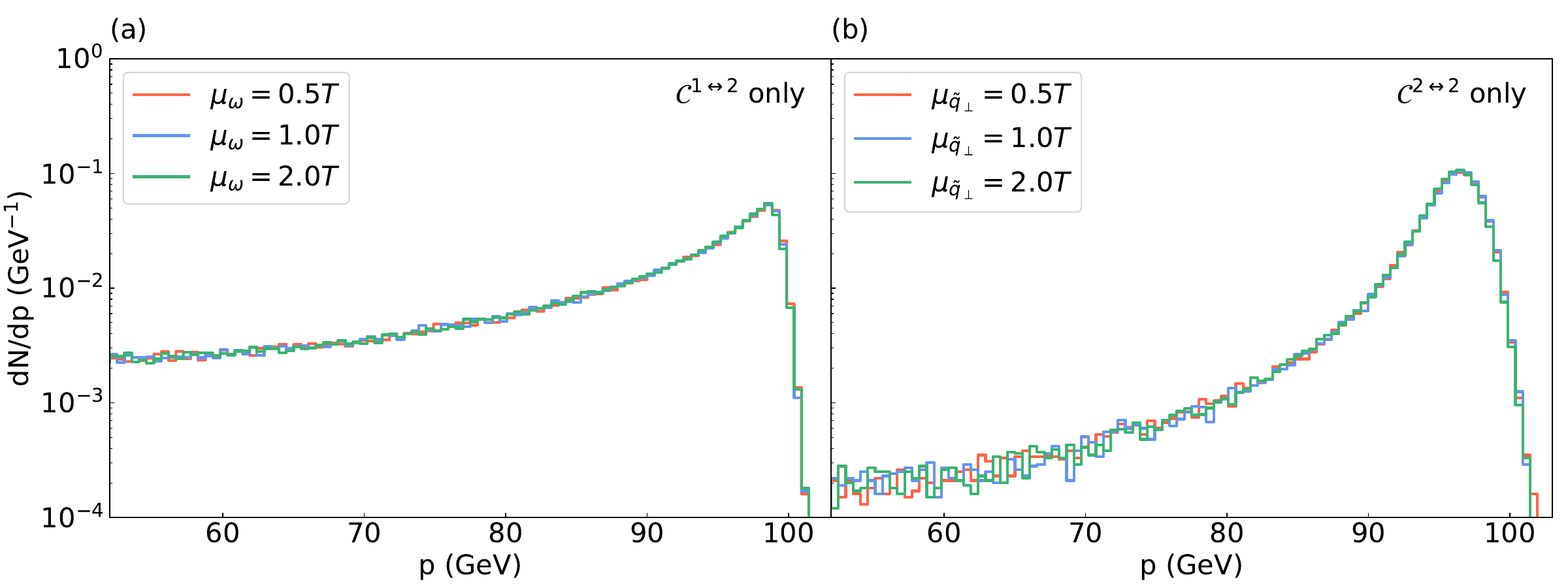}
    \caption{The energy distribution of a $100~\textrm{GeV}$ gluon propagating through $300~\textrm{MeV}$ QGP medium ($N_f=3$) at $\alpha_s = 0.005$ with different values of the cutoffs. The evolution time is $t=(0.3/\alpha_s )^2=3600$~fm. The subplot (a) only includes the $\mathcal{C}^{1\leftrightarrow 2}$ interactions and (b) only includes $\mathcal{C}^{2\leftrightarrow 2}$ interactions. In both cases the cutoff $\mu$ is varied: the soft interactions (those with momentum transfer less than $\mu_\omega$ and $\mutildeqperp$ respectively) are treated with a Langevin process, while the rest of the kinematic phase space is treated with rates. Results obtained when propagating an energetic light quark instead of a gluon can be found in Appendix~\ref{app:quark_energy_distribution}. }
    \label{fig:cutoff_dependence_small_coupling}
\end{figure*}

Building on the validation from the previous section,
we can combine our approaches for the hard and soft interactions to implement the entire hard-soft factorized parton energy loss model described in Section~\ref{sec:factorized_model}. Remember that in the following, we use vacuum matrix elements for $\mathcal{C}^{2\leftrightarrow 2}_{\textrm{hard}}$, since the screening effects are encoded in the drag and diffusion coefficients of soft interactions. We also extend this test to a full quark-gluon plasma, with $N_f=3$. We use once again $\alpha_s = 0.005$ ($g\approx 0.25$), with a propagation time of $t=(0.3/\alpha_s)^2=3600$~fm in a $T=300$~MeV plasma.

As summarized by Eq. (\ref{reformulation}), the hard or soft processes alone are dependent on the cutoff, but their cutoff dependence cancels out when combined. 
We confirm that, for both the inelastic and elastic cases, the single parton energy distribution is independent on the hard-soft cutoffs at small coupling
in Fig. \ref{fig:cutoff_dependence_small_coupling}, given a sufficiently long evolution time. These results are consistent with those obtained in the previous section. 

\section{Hard-soft factorization of parton energy loss beyond weak-coupling}
\label{sec:large_coupling}

Soft interactions between an energetic parton and a deconfined plasma are likely non-perturbative. Evaluating this non-perturbative rate from first principles is an ongoing challenge. In this section, we estimate this non-perturbative rate using a typical approach in the heavy ion literature: we use the perturbative rate and extrapolate it to large coupling. 

Recall that we do not use a running coupling in this work. As such, we use the same value of $\alpha_s$ for soft and hard interactions, with the understanding that the future introduction of a running coupling will indeed lead to smaller values of $\alpha_s$ for hard interactions, as assumed in this work.

As discussed in Section~\ref{sec:hard_soft_at_weak_coupling:soft_interactions:scale}, soft interactions can always be described stochastically, if propagation in the medium is sufficiently long. We quantified this duration as $t \gg \left<\omega^3\right>^2 / \left<\omega^2\right>^3$,
or $S \ll 1$ as defined in Eq.~(\ref{eq:scale}), with $\left<\omega^n\right>$ given by \eqref{eq:rate_moment}. We emphasize once again that Eq.~(\ref{eq:scale}) is general, and not limited to the perturbative regime.

We can use inelastic interactions
to get an estimate of the length of the medium required to describe soft interactions stochastically.
When extrapolating the weakly-coupled inelastic rate to large coupling, the $\omega$-dependence of the rate remains the same. This means that, within this approximation, the analytical expression for $\mathcal{S}$ --- Eq.~(\ref{eq:scale_inel}) --- remains the same.
Consequently, Eq.~(\ref{eq:time_langevin_inel}) remains the same as well, and it states that a stochastic description of inelastic interactions with $\omega < T$ requires a time $t \gg 1/[g^4 T]$. For temperatures of a few hundred MeV and a coupling $g\sim 1-2$ encountered in heavy ion collisions,  $1/[g^4 T] < 1$~fm. Under this estimate, it would be reasonable to describe stochastically soft interactions with $\mu \lesssim T$ occurring in a heavy ion collision. 

Note that the above conclusion is based on the estimate of the soft inelastic rate discussed above; should the non-perturbative rate differ significantly from it, it could lead to change the range of applicability of the Langevin equation. However, we do believe that the above estimates --- based on extrapolations of the weakly-coupled rates to strong coupling --- are encouraging.

In what follows, we use $\alpha_s=0.3$ ($g \approx 2$), and first compare a stochastic and a microscopic description of parton energy loss for soft interactions. We use a plasma of length $1$~fm and temperature $T=300$~MeV.

Note that, when the coupling is large, the analytical diffusion coefficients computed perturbatively are not necessarily consistent with numerical values obtained by direct integration of the rates (see Fig.~\ref{fig:qhat_coupling} and surrounding discussion). For what follows, we use the numerical diffusion coefficients in the Langevin part of the hard-soft factorized model.
\subsection{Comparison between diffusion process and collision rate}
\label{sec:comparison_in_large_coupling}

\begin{figure*}
    \centering
    \includegraphics[width=2\columnwidth]{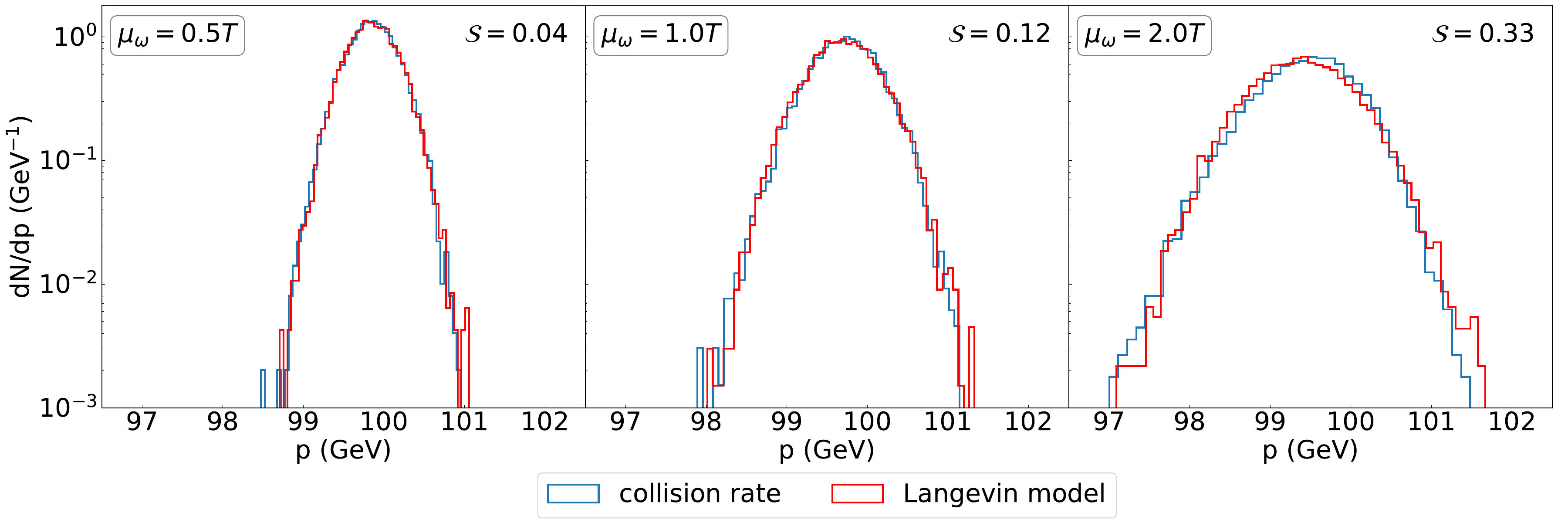}
    \caption{The energy distribution of a $100$~GeV gluon propagating through a $300$~MeV pure glue medium ($N_f = 0$) at $\alpha_s = 0.3$. The evolution time is $t = (0.3/\alpha_s )^2=1$~fm. Only soft $1\leftrightarrow 2$ interactions with $\omega < \mu_\omega$ are allowed. Compare with the weak-coupling result from Fig.~\ref{fig:diffusion_collision_rate_small_coupling:inel}.}
    \label{fig:inel_diffusion_collision_rate_large_coupling}
\end{figure*}

\begin{figure*}
    \includegraphics[width=2\columnwidth]{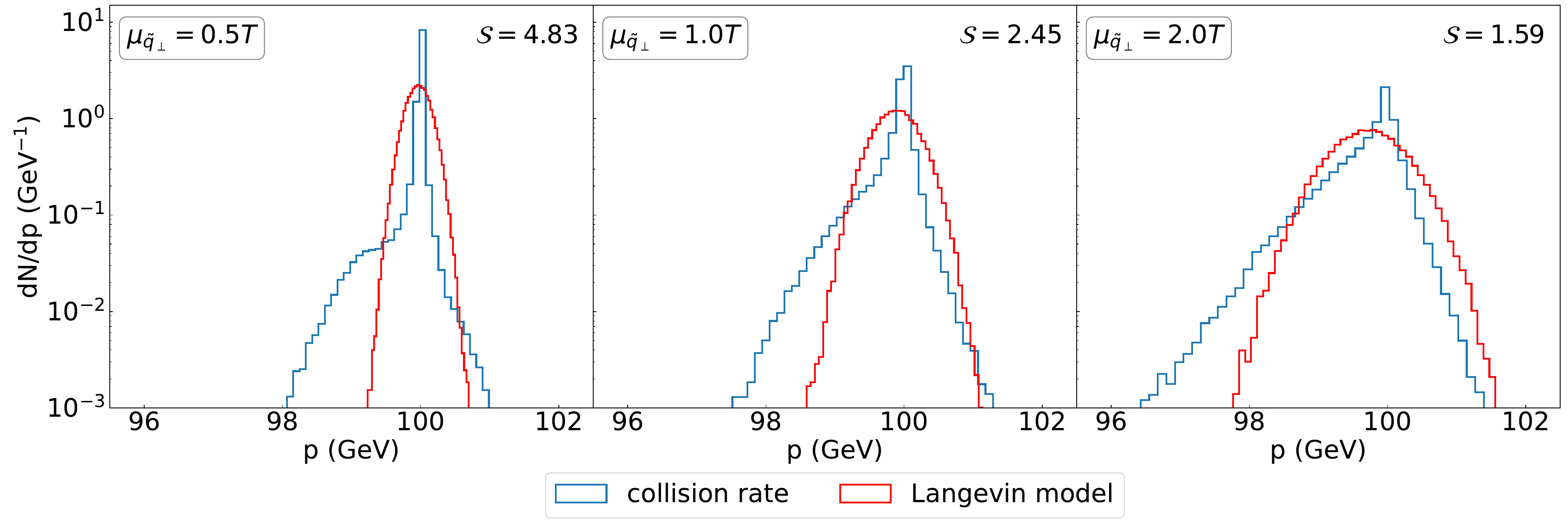}
    \onecolumngrid
    \caption{The energy distribution resulting from a $100$~GeV gluon propagating through a $300$~MeV pure glue medium ($N_f = 0$) at $\alpha_s = 0.3$. The evolution time is $t = (0.3/\alpha_s )^2=1$~fm. Only soft $2\leftrightarrow 2$ interactions with $\omega < \Lambda$ and $\tilde{q}_\perp < \mu_{\tilde{q}_\perp}$ are allowed. We choose $\Lambda=\min(p_{\textrm{cut}}, \sqrt{3p_0T})$. Compare with the weak-coupling result from Fig.~\ref{fig:diffusion_collision_rate_small_coupling:elastic}.}
    \label{fig:elas_diffusion_collision_rate_large_coupling}
\end{figure*}

\begin{figure*}
    \centering
    \includegraphics[width=2\columnwidth]{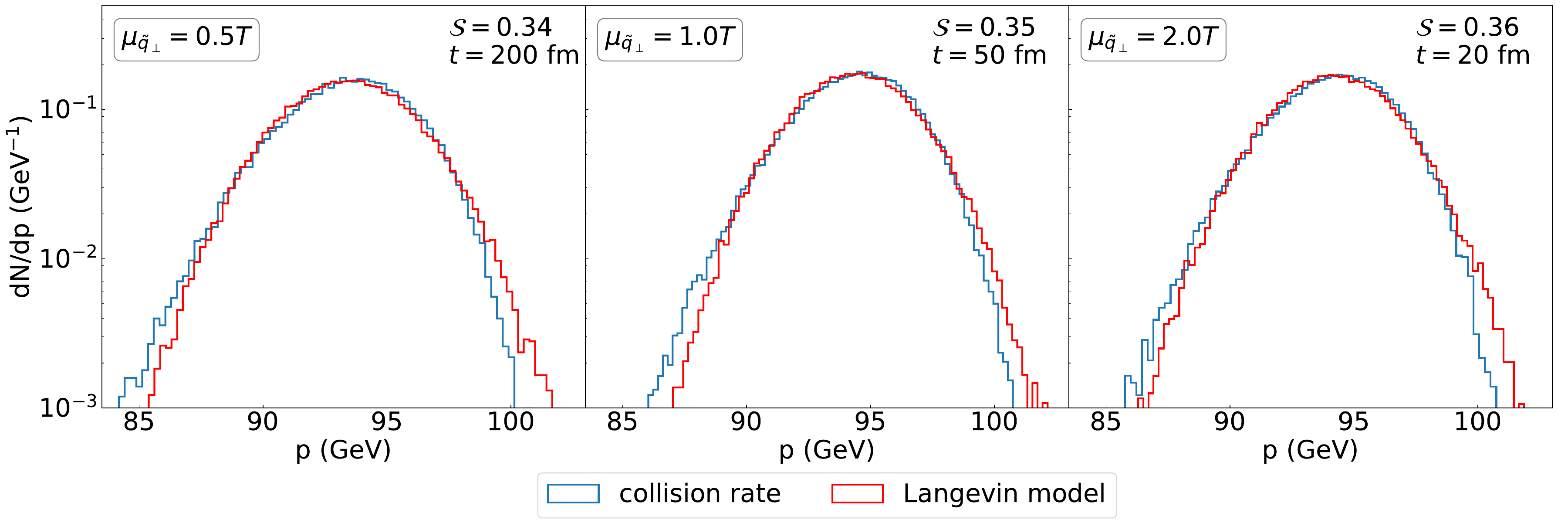}
    \caption{The energy distribution resulting from a $100$~GeV gluon propagating through a $300$~MeV pure glue medium ($N_f = 0$) at $\alpha_s = 0.3$. The evolution time is 200, 50, 20 fm for $\mu_{\tilde{q}_\perp}/T=0.5, 1, 2$; the times were chosen to obtain similarly small values of the skewness parameter $\mathcal{S}$ (\eqref{eq:scale}). Only soft $2\leftrightarrow 2$ interactions with $\omega < \Lambda$ and $\tilde{q}_\perp < \mu_{\tilde{q}_\perp}$ are allowed. We choose $\Lambda=\min(p_{\textrm{cut}}, \sqrt{3p_0T})$. }
    \label{fig:diffusion_collision_rate_large_coupling_large_t}
\end{figure*}

As in the weak-coupling case (Section~\ref{sec:weak_coupling_diffusion_vs_rate}), we perform this section's test in the pure glue limit ($N_f=0$). 

We first study the inelastic interactions, and as discussed above, we expect inelastic interactions softer than $\sim T$ to be describable by the Langevin equation in a $1$~fm brick. We show this explicitly in Fig.~\ref{fig:inel_diffusion_collision_rate_large_coupling}.
We show calculations for three different cutoffs $\mu_\omega$,
and we plot the results for the scale $\mathcal{S}$ from \eqref{eq:scale}.\footnote{We verified that the result from \eqref{eq:scale} is close to that of \eqref{eq:scale_inel}. The values we quote are from \eqref{eq:scale}} As expected, agreement between the Langevin approach and the microscopic collision rate approach are best when $\mathcal{S}\ll 1$. In the current setting, agreement is still good for $\mu_\omega = 2 T$, for which $\mathcal{S}=0.33$. This is encouraging evidence that the effect of non-perturbative inelastic interactions ($\mathcal{C}^{1\leftrightarrow 2}_{\textrm{soft}}$) can be treated stochastically in phenomenological applications such as heavy ion collisions.

The equivalent result for soft elastic interactions ($\mathcal{C}^{2\leftrightarrow 2}_{\textrm{soft}}$) is shown in Fig.~\ref{fig:elas_diffusion_collision_rate_large_coupling}. The result is very different. On one hand, the mean energy and width of the parton distribution described with the Langevin equation is almost identical to that described with collision rates. %
However their shape are different, especially at smaller values of the cutoffs $\mu_{\tilde{q}_\perp}$. 
Agreement between the two approaches is improved when the cutoff is larger. This is also reflected in the values of the scale $\mathcal{S}$, evaluated numerically with Eq.~(\ref{eq:scale}), which decreases with increasing $\mu_{\tilde{q}_\perp}$ (see Fig. \ref{fig:elastic_scale_num}).
This is different from what was observed (i) in the inelastic case (see Fig. \ref{fig:diffusion_collision_rate_small_coupling:inel}, \ref{fig:inel_diffusion_collision_rate_large_coupling}), and (ii) in the elastic case at weak coupling (see Fig. \ref{fig:diffusion_collision_rate_small_coupling:elastic}): both cases preferred smaller values of the cutoff.
Yet this result is fully consistent with our discussion in Section~\ref{sec:elastic_scale} of the scale $\mathcal{S}$ for the elastic rate: it is purely a consequence of the $\omega$-dependence of the elastic rate.
We verified in Fig.~\ref{fig:diffusion_collision_rate_large_coupling_large_t} that longer evolution times do lead to better agreement between the Langevin and the collision rate descriptions, reflected in smaller values of the scale $\mathcal{S}$.
Our tentative conclusion is that soft elastic collision may be more difficult to describe stochastically; it is possible that one needs a larger cutoff $\mu_{\tilde{q}_\perp}$ to describe these elastic interaction stochastically, although more studies will be necessary to confirm this conclusion. Note, however, that observables which are mainly sensitive to the average energy loss and the width of the parton distribution may tolerate a wider range of soft interactions being described with the Langevin approach. 

More generally, it is clear that the choice of cutoff is very important in stochastic descriptions: careful choices of cutoffs can broaden significantly the range of applicability of the factorized approach presented in this work. Importantly, the cutoff choice should be chosen based on the expected relative size of the third and second moments of the energy loss rates.

\subsection{Parton energy loss at large coupling in a static medium}
\label{sec:parton_evolution_large}

\begin{figure*}
    \centering
    \includegraphics[width=2\columnwidth]{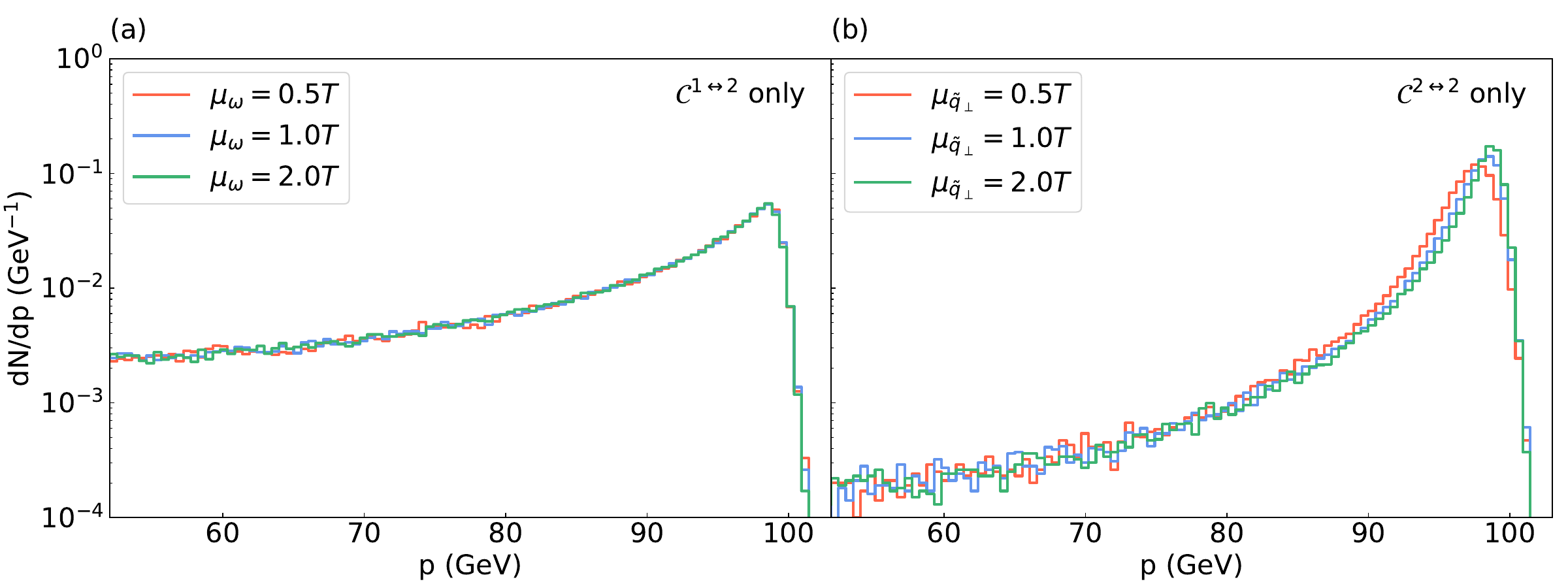}
    \caption{The energy distribution of a $100~\textrm{GeV}$ gluon propagating through $300~\textrm{MeV}$ QGP medium ($N_f=3$) at $\alpha_s = 0.3$ with different values of the cutoff. The evolution time is $t=(0.3/\alpha_s)^2=1~\textrm{fm}$. The subplot (a)  only includes $\mathcal{C}^{1\leftrightarrow 2}$ interactions and (b) only includes $\mathcal{C}^{2\leftrightarrow 2}$ interactions. See the weakly-coupled results in Fig.~\ref{fig:cutoff_dependence_small_coupling} for comparison and additional explanations. Results obtained when propagating an energetic light quark instead of a gluon can be found in Appendix~\ref{app:quark_energy_distribution}.}
    \label{fig:cutoff_dependence_large_coupling}
\end{figure*}

To close this section, we quantify the cutoff dependence of a $100~\textrm{GeV}$ parton propagating for $1$~fm in a $300$~MeV brick of plasma, with $\alpha_s = 0.3$. This ``brick'' is the same as in the previous section. The soft interactions are described with the Langevin equation, and hard interactions are included as in the full implementation of the hard-soft energy loss model (Section~\ref{sec:factorized_model}). We use $N_f = 3$ in this test. 

We plot the energy distributions with different values of the cutoff in Fig.~\ref{fig:cutoff_dependence_large_coupling}.
In this larger coupling regime, as expected from the results of the previous section, inelastic interactions ($\mathcal{C}^{1\leftrightarrow 2}$) are independent of the cutoff (panel (a)). 
For the elastic case ($\mathcal{C}^{2\leftrightarrow 2}$), the energy distributions with different values of the cutoff are slightly different in the large energy region, although the long tail of the distribution is not affected (panel (b)).

Note that we also performed a cutoff dependence test on the cutoff $\Lambda$ for $2\leftrightarrow 2$ interactions. We found the energy distribution of a parton propagating in a static medium to be independent of the choice of $\Lambda$, as expected. The result and further discussion can be found in \app{app:lambda_dependence}.

\section{Application: Energy and fermion-number cascade}

In this section, we use the hard-soft factorized model to study the energy and fermion-number cascade resulting from inelastic interactions between an energetic parton and a thermal medium. This section thus focuses on $\mathcal{C}^{1\leftrightarrow 2}$ (Fig. \ref{fig:reformulation}-b) in the hard-soft factorized model; both the hard and soft inelastic interactions are included, with the soft inelastic interactions modeled by the Langevin evolution.  The collision kernel $\mathcal{C}^{2\leftrightarrow 2}$ is switched off for this section. %

\subsection{Energy cascade of hard gluons}
\label{sec:energy_cascade}

\begin{figure}
    \centering
    \includegraphics[width=\linewidth]{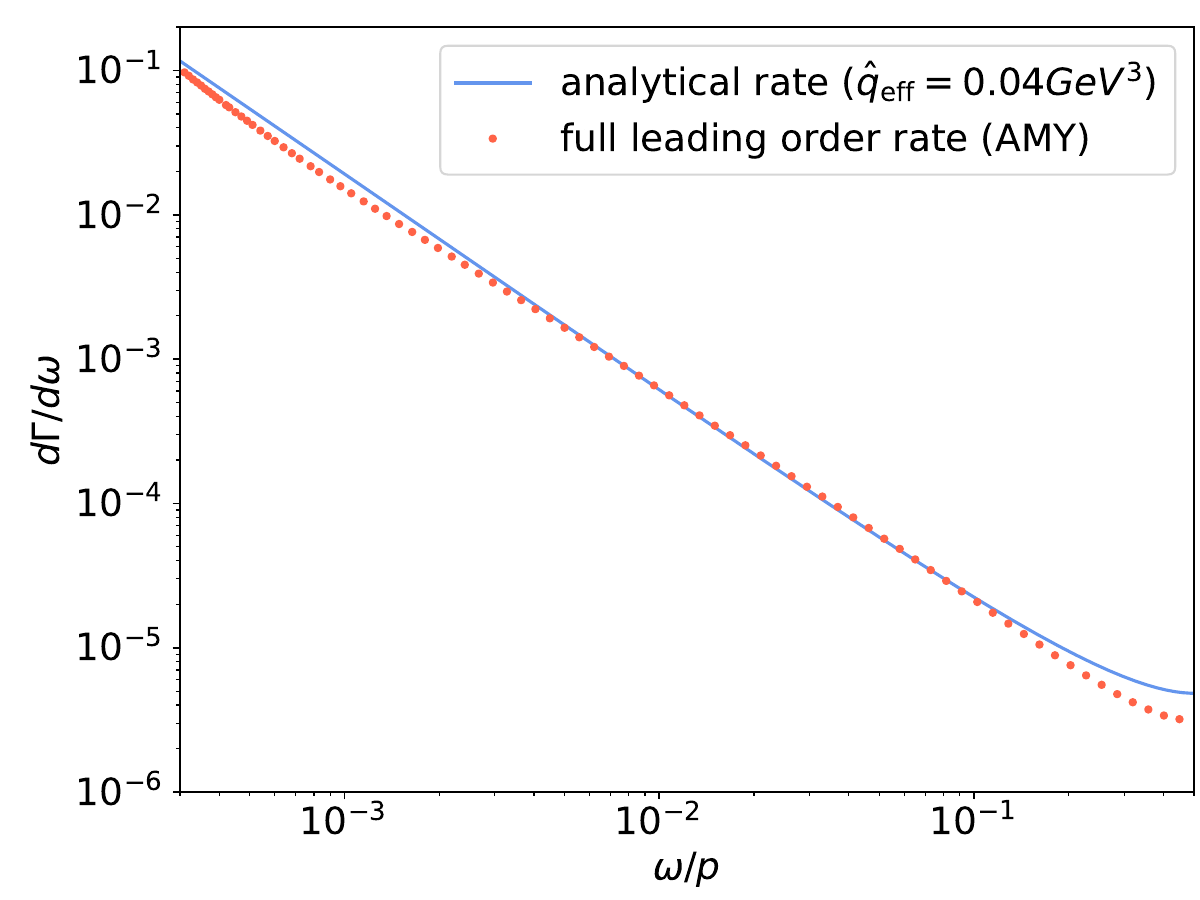}
    \caption{Comparison between the full leading-order inelastic rate and the deep-LPM regime approximation of the rate from \eqref{eq:branching_rate} with $\hat{q}_{\rm eff}=0.04$~GeV$^3$ for $N_f=0$ ($p=1$~TeV, $T=300$~MeV and $\alpha_s=0.1$). %
    }
    \label{fig:rate_nf0}
\end{figure}

\begin{figure}
    \centering
    \includegraphics[width=\columnwidth]{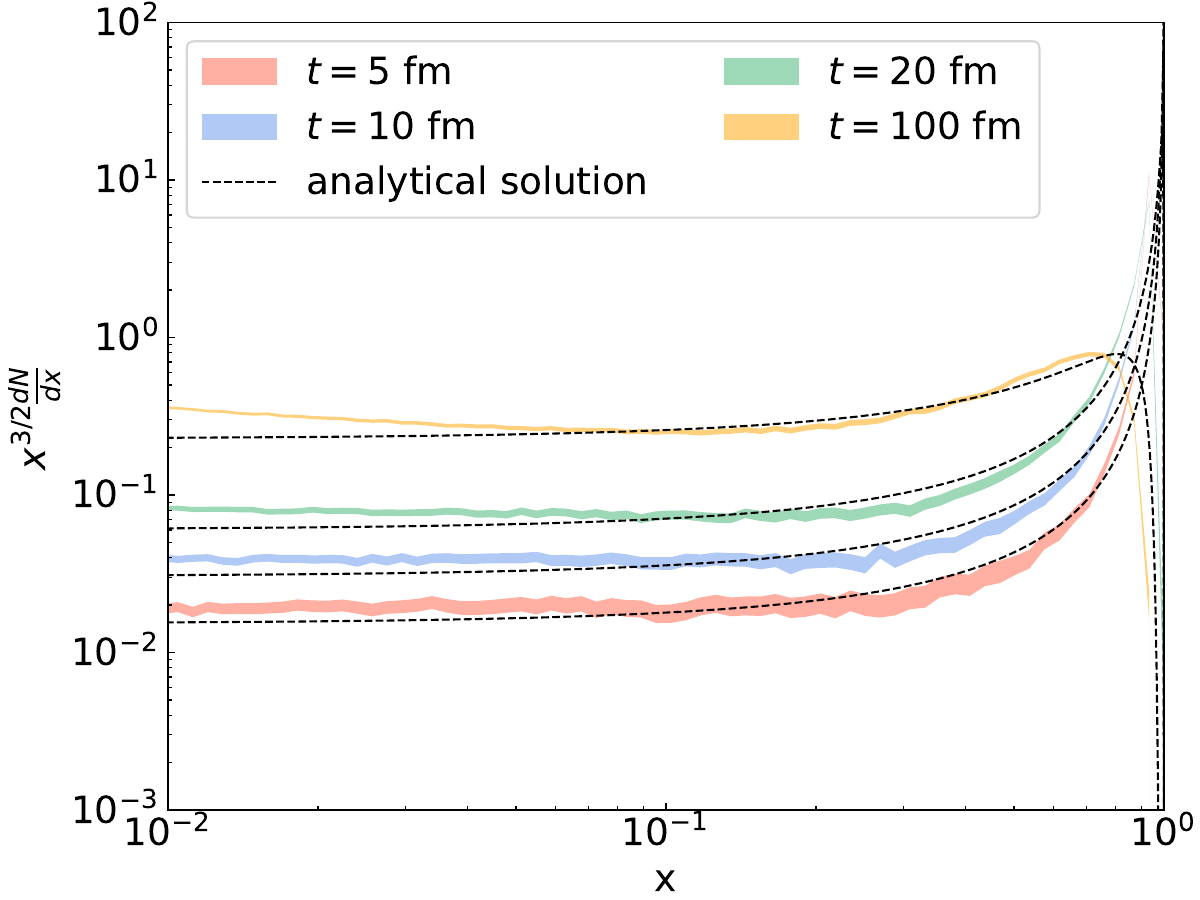}
    \caption{A comparison of the current numerical implementation of QCD kinetics and the analytical approximation of Ref.~\cite{Blaizot:2013hx} for the energy cascade in the pure glue medium for different evolution times. The analytical solution is denoted by the dotted curve. 
    In this test, we only include inelastic $1\leftrightarrow 2$ processes.
    We use $N_f=0$, $\alpha_s=0.1$, $T=300$~MeV and $p_0=1000$~GeV. }
    \label{fig:energy_scaling}
\end{figure}

When a gluon propagates through a thermal QCD medium, successive medium-induced inelastic radiations result in a gluon cascade. An analytical approximation for the gluon cascade was introduced in Refs.~\cite{Blaizot:2013hx,Blaizot:2015jea}; it was argued that the successive medium-induced quasi-democratic emissions lead to the accumulation of gluons at zero energy and cause a power-law scaling in the small energy region.  We will study this scaling in this section~\cite{[{A pedagogical introduction relating  the AMY kinetic equations to the turbulent cascade presented  here is given in: }]Schlichting:2019abc}.

At leading order, the successive radiations can be assumed to be independent~\cite{Blaizot:2013vha}. In the deep LPM region, where the time scale of the radiation process is much larger than the mean free path between multiple scatterings, the rate per unit time of a gluon with energy $p$ splitting into two gluons with energy fractions $z$ and $1-z$   can be approximated 
as\footnote{Accounting for the identical particles in  the final state, the total rate is $\int_0^{1/2} d\Gamma/dz \,dz$. }~\cite{Blaizot:2013vha, Arnold:2008vd,PhysRevD.79.065025}
\begin{equation}
   \left.\frac{d\Gamma}{dz}\right|_{g\leftrightarrow gg} =\frac{\alpha_s N_c}{\pi}\frac{1}{[z(1-z)]^{3/2}}\sqrt{\frac{\hat{q}_{\rm eff}}{p}}  \, .
\label{eq:branching_rate}
\end{equation}
Here $\hat{q}_{\rm eff}$ is the average transverse momentum broadening of the radiated gluon, and $z=\omega/p$ with $\omega$ the energy of the radiated gluon. 
We have kept only the most singular parts of the splitting 
function at $z \sim 0$. %
We  will  treat $\hat q_{\rm eff}$ as a fit parameter, and then relate it to the parameter $\hat q_{\rm soft}^{2\leftrightarrow 2}$ in \eqref{eq:elas_qhat}.

The energy of the initial gluon is $p_0$, and we define $x \equiv \omega/p_0$.  The 
evolution of the gluon spectrum $D(x, \tau) = x(dN/dx)$ is governed by~\cite{Blaizot:2013hx,Schlichting:2019abc}
\begin{multline}
   \frac{\partial D(x, \tau)}{\partial \tau} = \int_0^{1} dz \frac{1}{[z(1-z)]^{3/2}} \\
    \times 
    \left[\sqrt{\frac{z}{x}}D\left(\frac{x}{z}, \tau\right)-\frac{z}{\sqrt{x}}D\left(x, \tau\right)\right], 
    \label{eq:FP_evolution}
 \end{multline}
where
\begin{equation}
   \tau \equiv \frac{\alpha_sN_c}{\pi}\sqrt{\frac{\hat{q}_{\rm eff}}{p_0}}t \, ,
\end{equation}
and $t$ is the evolution time of the gluon. 

The exact solution for \eqref{eq:FP_evolution} can be calculated via Laplace transform:
\begin{equation}
    D_0(x, \tau)=\frac{\tau}{\sqrt{x}(1-x)^{3/2}}e^{-\pi[\tau^2/(1-x)]}. 
    \label{eq:spectrum}
\end{equation}
As remarked in Ref.~\cite{Blaizot:2013hx}, this power-law gluon spectrum Eq. (\ref{eq:spectrum}) scales as $1/\sqrt{x}$ in the small-x region.  

In order to compare  with \eqref{eq:spectrum}, we first determine the approximate value of $\hat{q}_{\rm eff}$ to use in the simplified rate \eqref{eq:branching_rate}; this value also enters Eqs.~(\ref{eq:FP_evolution}-\ref{eq:spectrum}).
We fix $\hat{q}_{\rm eff}$ by comparing \eqref{eq:branching_rate} with the full leading-order inelastic rate, as shown in Fig.~\ref{fig:rate_nf0}. With parameters given in Fig.~\ref{fig:rate_nf0}, we find  $\hat{q}_{\rm eff}\simeq 0.04$~GeV$^3$ at $\omega/p\simeq 10^{-2}$. We will use this value of $\hat q_{\rm eff}$ in our analysis of the cascade below.

It should be emphasized that \eqref{eq:branching_rate} is an approximation to the full inelastic rates corresponding to \eqref{eq:1to2}. Indeed, a leading-log analysis of the full rates at small $z$ in the deep LPM regime shows that~\cite{Arnold:2008vd}:
\begin{equation}
   \hat q_{\rm eff} = \hat q_{\rm soft}^{2\leftrightarrow 2} (\mu_\perp^2) \, ,
   \label{eq:qeff_est}
\end{equation}
where  $\hat{q}_{\rm soft}^{2\leftrightarrow 2}$ is given in \eqref{eq:elas_qhat}, and  $\mu_\perp^2=C_0\sqrt{2 \omega \hat q_{\rm eff} }$ with $C_0 \sim 1$.
The cutoff $\mu_\perp^2$ scales with 
the accumulated
transverse momentum of the radiated gluon over its formation time. A next-to-leading logarithmic analysis fixes the coefficient $C_0$~\cite{Arnold:2008zu}: 
\begin{equation}
\mu_\perp^2 = C_0 \sqrt{2 \omega \hat{q}_{\rm eff}}\, ,   \qquad C_0 = 2e^{2 -\gamma_E + \pi/4} \, .
\label{eq:muperp_est}
\end{equation}

For $N_f=0$, $p=1$~TeV, $T=300$~MeV, $\alpha_s=0.1$ (same as in \Fig{fig:rate_nf0}), and using $\omega/p=10^{-2}$, we can solve Eqs.~(\ref{eq:qeff_est}-\ref{eq:muperp_est}) numerically. We find $\hat{q}_{\rm eff}\approx 0.052~$GeV$^3$, which as expected is close to the value we found in Fig.~\ref{fig:rate_nf0}.

We next perform the gluon cascade in a pure-glue medium ($N_f=0$) using the hard-soft factorized model, i.e we  include both soft inelastic interactions described with the Langevin equation, and rate-based hard inelastic interactions which dominate this test. 
In Fig. \ref{fig:energy_scaling}, we compare this numerical result calculated by the current model with the analytical spectrum in Eq. (\ref{eq:spectrum}). We find that the numerical solution for the medium-induced cascade is reasonably well described by the approximate analytic solution. In particular, the power 
law behavior, $dN/dx \propto x^{-3/2}$, is nicely captured by this solution.

\subsection{Fermion-number cascade of gluons and quarks}

\begin{figure}
    \centering
    \includegraphics[width=\columnwidth]{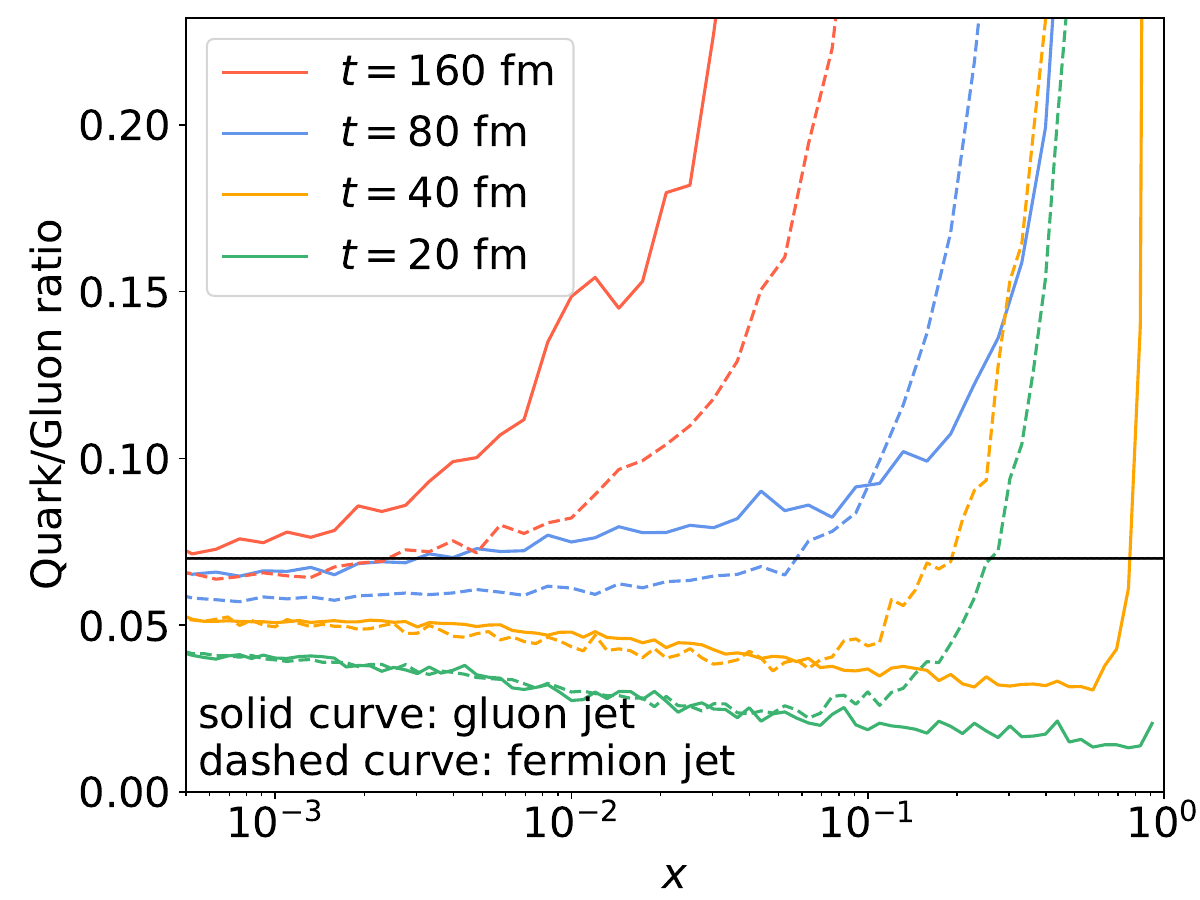}
    \caption{The fermion number cascade of the numerical implementation in the QGP medium with different evolution times. In this test, we only include inelastic interactions ($\mathcal{C}^{1\leftrightarrow 2}$). We use $N_f=3$, $\alpha_s=0.3$, $T=300$~MeV and $p_0=10$~TeV. The black horizontal line reflects the expected limiting value of $D_S/(2 N_f D_g) \approx 0.07$ (\eqref{eq:fermion_ratio}).}
    \label{fig:fermion_cascade}
\end{figure}

The fermion-number cascade was investigated in Ref.~\cite{Mehtar-Tani:2018zba}. Given the power-law scaling in the small energy region, at small $x$, we can write the power-law spectrum of quarks and gluons as
\begin{equation}
    \begin{split}
        D_g &\equiv x \frac{dN_g}{dx} = \frac{G}{\sqrt{x}},  \\
        D_s &\equiv \sum_{i = 1}^{N_F}\left(D_{q_i}+D_{\bar{q}_i}\right) = \frac{Q}{\sqrt{x}}. 
    \end{split}
\end{equation}

As derived in Ref.~\cite{Mehtar-Tani:2018zba}, the quark-to-gluon ratio of the soft radiated partons is determined by the transformation rate between gluons and fermions. We have
\begin{equation}
    \frac{Q}{2N_fG} = \frac{1}{2N_f}\frac{\int_0^1 dz z \mathcal{K}_{qg}(z)}{\int_0^1 dz z \mathcal{K}_{gq}(z)} \approx 0.07, 
    \label{eq:fermion_ratio}
\end{equation}
where $\mathcal{K}_{qg}$ is the splitting function of $g \rightarrow q\bar{q}$, and $\mathcal{K}_{gq}$ is the splitting function of $q \rightarrow gq$. 

To test the quark-to-gluon ratio in the hard-soft factorized model, we numerically simulate the evolution of a gluon or a quark propagating through a static QGP medium ($N_f=3$) using the full leading order inelastic rate. 
We perform the calculation for both an energetic gluon and an energetic light quark with an initial energy of $10$~TeV.
The result is shown in Fig. \ref{fig:fermion_cascade}; we find that it converges to the universal quark-to-gluon ratio when using the full QCD rates.

\section{Summary and outlook}

This work introduces a new formulation of parton energy loss where soft and hard interactions with the underlying plasma are factorized and treated separately.
The factorization is performed with cutoffs based on the momentum transfer of the interactions.
Rare hard interactions are considered as independent successive interactions, and solved with collision rates (Sections~\ref{sec:factorized_model:inelastic_hard} and \ref{sec:factorized_model:elastic_hard});
the larger momentum exchange with the medium make them more likely to be amenable to a perturbative description.
On the other hand, frequent soft interactions are treated stochastically using a Langevin evolution with drag and diffusion coefficients encoding the effect of these soft interactions (Section~\ref{sec:hard_soft_theory:reformulation:soft}); non-perturbative effects can thus be absorbed in these transport coefficients. 

Our numerical implementation of this model (Section~\ref{sec:tequila_weak_coupling}) shows that this factorization works well in the weakly-coupled regime where the theory was derived~\cite{Ghiglieri:2015ala}.
In fact, by revisiting the conditions under which the Langevin equation can describe the Boltzmann equation (Section~\ref{sec:hard_soft_at_weak_coupling:soft_interactions:scale}), we extended the region of phase space (``cutoffs'') that can be described stochastically. We used the dimensionless scale $\mathcal{S}$ (\eqref{eq:scale}) to quantify the length of a plasma necessary for soft collisions to be describable with the Langevin equation. Our numerical tests showed that this scale works very well in practice.

Because the scale $\mathcal{S}$ is a property of the Boltzmann equation and not a perturbative concept, we used it to extend our discussion of parton energy loss beyond the perturbative regime. We estimated that inelastic collisions resulting in parton energy loss of order $T$ could be described stochastically in a QCD plasma of size $\sim 1$~fm (Section~\ref{sec:large_coupling}). Given that inelastic interactions dominate parton energy loss for high-energy partons, this supports the applicability of the present energy loss model in heavy ion collisions.

This work paves the way to systematic phenomenological constraints on the soft transport coefficients of light partons.
The key strength of our approach is that perturbative parton energy loss calculations are still being used for harder interactions --- the regions of phase space where they are most likely to hold.
Conversely, the interactions most sensitive to non-perturbative effects --- soft interactions --- are encoded in simple transport coefficients which can be constrained by comparison with measurements.
A stochastic description of soft collisions can also be very efficient numerically, as a large number of soft interactions can be absorbed in the transport coefficients.
These phenomenologically-constrained transport coefficients can eventually be compared with  lattice results~(e.g. Ref.~\cite{Moore:2019lgw}).
A similar program is already being pursued for the energy loss of heavy quarks~\cite{Ke:2018tsh}; studies of light parton energy loss with a model that includes many features of soft-hard factorization, are also ongoing~\cite{Ke:2020clc}.

Future generalization of this framework includes improving the treatment of the radiation angle of collinear radiation, and the inclusion of a running coupling constant and of next-to-leading order effects; these additions will increase the type of observables that can be studied with this model. The inclusion of finite-size effects in this formalism will also be an important addition. These additions will be able to build on Ref.~\cite{Ke:2020clc} and other works.

\begin{acknowledgments}
We thank Weiyao Ke for his invaluable help in the early stage of this project, and Jacopo Ghiglieri for generously sharing notes on the splitting approximation that formed the basis of this work's discussion on the topic.
We thank Sangyong Jeon, Chanwook Park, Abhijit Majumder and the other members of the JETSCAPE Collaboration for discussions regarding MARTINI and MATTER, and for their support with the JETSCAPE framework.
We thank Yingru Xu for valuable discussions regarding the Langevin equation and its numerical implementation.
This work was supported by the U.S. Department of Energy Grant no. DE-FG02-05ER41367 (SAB, JFP and TD) and DE-FG-02-08ER41450 (DT).
TD is also supported by NSF grant OAC-1550225. This research used resources of the National Energy Research Scientific Computing Center (NERSC), a U.S. Department of Energy Office of Science User Facility operated under Contract No. DE-AC02-05CH11231.
\end{acknowledgments}

\appendix

\section{Inelastic rate at low $\omega$}

\label{appendix:inel_low_omega}

At leading order, the differential rate of the $1\leftrightarrow 2$ process can be expressed using AMY's rate~\cite{Jeon:2003gi, CaronHuot:2010bp}: 
\begin{equation}
\begin{split}
\left.\frac{d\Gamma(p, \omega)}{d\omega}\right|^{1\leftrightarrow 2} &= \frac{g^2}{16\pi p^3 \omega^2 (p-\omega)^2}\left[1\pm n(\omega)\right]\left[1 \pm n(p-\omega)\right] \\
&\times P_{bc}^{a}(z) \int \frac{d^2h}{(2\pi)^2}2\mathbf{h}\cdot \textrm{Re} \mathbf{F}(\mathbf{h}, p, \omega) 
\end{split}
\end{equation}
where $z = \omega/p$ and $P_{bc}^{a}(z)$ are the Dokshitzer-Gribov-Lipatov-Altarelli-Parisi (DGLAP) splitting kernels of the radiation $a\rightarrow bc$, 
\begin{equation}
    P_{bc}^{a}(z) = \left\{
    \begin{aligned}
    & C_F\frac{1+(1-z)^2}{z}, \quad &q \rightarrow gq \\
    & C_A \frac{1+z^4+(1-z)^4}{z(1-z)}, \quad &g \rightarrow gg \\
    & \frac{d_FC_F}{d_A}\left[z^2+(1-z)^2\right], \quad &g \rightarrow q\bar{q}
    \end{aligned}
    \right. .
\end{equation}
Very soft interactions ($\omega \ll T$) are dominated by gluon radiations, i.e. $g \leftrightarrow gg$, $q\leftrightarrow gq$ with a soft final state gluon (see footnote \ref{footnote:gluons_low_omega}). 
In this case ($\omega \ll T \ll p$), AMY's integral is symmetric and can be expanded in terms of the radiated energy $\omega$~\cite{Ghiglieri:2015ala}: 
\begin{equation}
\begin{split}
&\left. \int \frac{d^2h}{(2\pi)^2}2\mathbf{h}\cdot \mathrm{Re} \mathbf{F}(\mathbf{h}, p, \omega)\right|_{\textrm{soft gluon}} = 8p^6 C_A z^2(1-2z) \int\frac{d^2q_\perp}{(2\pi)^2} \\
& \times \int\frac{d^2k_\perp}{(2\pi)^2}\frac{\mathcal{C}_F(k_\perp)}{C_F}\left[\frac{\mathbf{q}_\perp}{q_\perp^2+M_\infty^2}-\frac{\mathbf{q}_\perp+\mathbf{k}_\perp}{(\mathbf{k}_\perp+\mathbf{q}_\perp)^2+M_\infty^2}\right]^2, 
\end{split}
\label{eq:soft_radiation}
\end{equation}
where the collision kernel is
\begin{equation}
    \frac{\mathcal{C}_F(k_\perp)}{C_F} = \frac{g^2Tm_D^2}{k_\perp^2(k_\perp^2+m_D^2)}. 
\end{equation}
We define the integral in Eq. (\ref{eq:soft_radiation}) as 
\begin{equation}
\begin{split}
    \mathcal{I} = &\int \frac{d^2q_\perp}{(2\pi)^2}\int\frac{d^2k_\perp}{(2\pi)^2}\frac{\mathcal{C}_F(k_\perp)}{C_F} \\
    & \times \left[\frac{\mathbf{q}_\perp}{q_\perp^2+M_\infty^2}-\frac{\mathbf{q}_\perp+\mathbf{k}_\perp}{(\mathbf{k}_\perp+\mathbf{q}_\perp)^2+M_\infty^2}\right]^2. 
\end{split}
\label{eq:integral}
\end{equation}
Combining the factors, we have
\begin{equation}
\begin{split}
   \mathcal{I} &= g^2Tm_D^2\int \frac{dq_\perp^2}{4\pi}\int\frac{dk_\perp^2}{4\pi}\frac{1}{k_\perp^2(k_\perp^2+2M_\infty^2)} \\
   &\int \frac{d\phi_q}{2\pi}\int \frac{d\phi_{kq}}{2\pi}\left[\frac{\mathbf{q}_\perp}{q_\perp^2+M_\infty^2}-\frac{\mathbf{q}_\perp+\mathbf{k}_\perp}{(\mathbf{k}_\perp+\mathbf{q}_\perp)^2+M_\infty^2}\right]^2
\end{split}
\end{equation}
By rescaling all the dimensional quantities by $M_\infty$, the integral $\mathcal{I}$ can be calculated as 
\begin{equation}
\begin{split}
& \mathcal{I} = g^2T\frac{m_D^2}{M_\infty^2}\int \frac{d\hat{q}_\perp^2}{4\pi}\int\frac{d\hat{k}_\perp^2}{4\pi}\frac{1}{\hat{k}_\perp^2(\hat{k}_\perp^2+2)} \\
&\times \int \frac{d\phi_q}{2\pi}\int \frac{d\phi_{kq}}{2\pi}\left[\frac{\hat{q}_\perp}{q_\perp^2+1}-\frac{\hat{q}_\perp+\hat{k}_\perp}{(\hat{k}_\perp+\hat{q}_\perp)^2+1}\right]^2\\
& = \frac{2-\log(2)}{8\pi^2}g^2T. 
\end{split}
\end{equation}

We therefore have an analytical approximation of the soft gluon radiation rates: 
\begin{equation}
\begin{split}
&\left.\frac{d\Gamma(p, \omega)}{d\omega}\right|_{\textrm{soft gluon}}^{1\leftrightarrow 2} = \frac{\left[2-\log(2)\right]g^4 C_AT}{16\pi^3 p} \\
&\times \left[1 \pm n(\omega)\right]\left[1 \pm n(p-\omega)\right] \frac{1-2z}{(1-z)^2}P_{bc}^{a}(z). 
\end{split}
\label{eq:soft_differential_rate_ggg}
\end{equation}

\begin{figure}
    \centering
    \includegraphics[width=\linewidth]{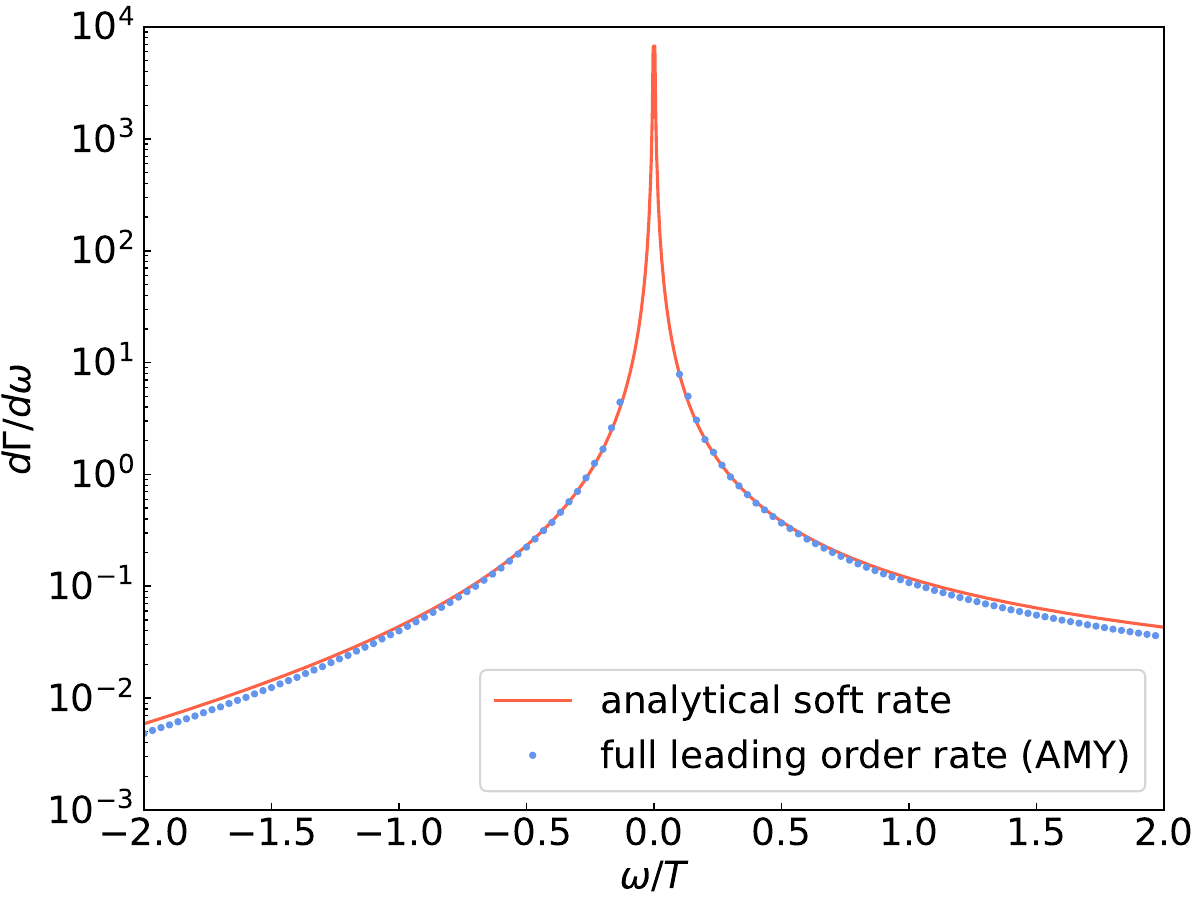}
    \caption{Comparison of the $g\leftrightarrow gg$ collision rate between the soft analytical expression (\eqref{eq:soft_differential_rate_ggg}) and the full leading order rate. We used $N_f = 3$, $\alpha_s=0.3$, $T=300$~MeV and $p_0=100$~GeV. }
    \label{fig:differential_rate_ggg}
\end{figure}

For the tests in Sections~\ref{sec:tequila_weak_coupling} and \ref{sec:large_coupling}, we use this soft limit of differential rate when $|\omega| \leq 0.2T$. In Fig. (\ref{fig:differential_rate_ggg}), we compare this soft limit with AMY's full rate for $g\leftrightarrow gg$, and they agree well in the soft $\omega$ region. 

With the soft radiation assumption $\omega \ll T \ll p$, we can simplify Equation (\ref{eq:soft_differential_rate_ggg}) by neglecting the terms suppressed by $\omega/T$ and $\omega/p$ 
\begin{equation}
\left.\frac{d\Gamma(p, \omega)}{d\omega}\right|_{\textrm{soft gluon}}^{1\leftrightarrow 2} \approx \frac{\left[2-\log(2)\right]g^4 C_AC_RT^2}{8\pi^3 \omega^2}. 
\label{eq:inel_soft_rate}
\end{equation}

Using the above expressions, we can calculate the perturbative $\hat{q}_{\textrm{L, soft}}^{1\leftrightarrow 2}$. 
We find that the longitudinal momentum broadening of soft $1\leftrightarrow 2$ is 
\begin{equation}
\begin{split}
    \hat{q}_{\textrm{L, soft}}^{1\leftrightarrow 2} &= \int_{-\mu_\omega}^{\mu_\omega} d\omega \omega^2 \left.\frac{d\Gamma(p, \omega)}{d\omega}\right|_{\textrm{soft gluon}}^{1\leftrightarrow 2} \\
    & = \frac{\left[2-\log(2)\right]}{4\pi^3}g^4C_RC_AT^2\mu_\omega. 
\end{split}
\end{equation}

\section{Energy loss rate for hard $2\leftrightarrow 2$ interactions}
\label{app:remaining}
The differential energy loss rate of a hard $2\leftrightarrow 2$ interaction is calculated using the vacuum matrix elements: 
\begin{equation}
   \frac{d^2\Gamma^{ab\leftrightarrow cd}_{\textrm{vac}}}{d\omega d\tilde{q}_\perp} = \int_{\frac{q-\omega}{2}}^{\infty}dk \int_0^{2\pi}\frac{d\phi}{2\pi}\frac{1}{4(2\pi)^3}\frac{\tilde{q}_\perp}{q}\frac{Q^{ab}_{cd}(p, k, \omega, \tilde{q}_\perp, \phi)}{4p^2},
    \label{eq:differential_rate}
\end{equation}
with 
\begin{equation}
        Q^{ab}_{cd}(p, k, \omega, \tilde{q}_\perp, \phi) =  \frac{1}{\nu_a}|\mathcal{M}^{ab}_{cd}|^2\left[n_b(k)(1\pm n_d(k+\omega))\right], 
        \label{eq:large_angle_total}
\end{equation}
where $\nu_a = 2d_a$ is the degeneracy of the particle $a$, $\tilde{q}_\perp = \sqrt{q^2-\omega^2}$, $\mathcal{M}^{ab}_{cd}$ is the matrix element of a vacuum $2\leftrightarrow 2$ interaction as a hard parton $a$ interacting with a thermal particle $b$ and transforms into particles $c$ and $d$. The expression of $\mathcal{M}^{ab}_{cd}$ can be found in Table II in \cite{Arnold:2002zm}. 

The collision kernel of the $2\leftrightarrow 2$ large-angle interactions is 
\begin{equation}
    \mathcal{C}^{2\leftrightarrow2}_{\textrm{large-angle}} = \sum_{bcd}\int_{-\infty}^{\Lambda} d\omega \int_{\mu_{\tilde{q}_\perp}}^\infty d\tilde{q}_\perp \frac{d^2\Gamma^{ab\leftrightarrow cd}_{\textrm{vac}}}{d\omega d\tilde{q}_\perp}. 
    \label{eq:total_kernel}
\end{equation}

In Eq. (\ref{eq:total_kernel}), if outgoing particles $c$ and $d$ are identical species, a symmetry factor of $\frac{1}{2}$ should be included. However, this factor of $\frac{1}{2}$ is canceled out to incorporate the interactions with $p- \Lambda < \omega < p$, since symmetric $2\leftrightarrow 2$ interactions with $p- \Lambda < \omega < p$ are equivalent to interactions with $\omega < \Lambda$. For $c$ and $d$ being distinct species, a factor of $\frac{1}{2}$ is also necessary to cancel the double-count of the final states in $\sum_{cd}$. We eliminate this factor by constraining that the energy of particle $c$ is larger than particle $d$. These asymmetric interactions with $\omega < \Lambda$ and $p- \Lambda < \omega < p$ are treated separately. %

In Eq. (\ref{eq:differential_rate}), the expression of $Q^{ab}_{cd}(p, k, \omega, \tilde{q}_\perp, \phi)$ is dependent on the types of particles $a$, $b$, $c$, and $d$. We summarize them using Mandelstam variables ($s$, $t$, $u$), Casimir factors ($C_A$, $C_F$), and color degrees of freedom ($d_F$, $d_A$) as follows, where $C_A=3$, $C_F=4/3$, $d_A = N_c^2-1$, $d_F=N_c$. We summarize the expression of $Q^{ab}_{cd}(p, k, \omega, \tilde{q}_\perp, \phi)$ for different interactions in Table \ref{tab:matrix_elements}. %

\begin{widetext}
\bgroup
\def\arraystretch{2}
\begin{center}
\begin{table}
\begin{tabular}{ | c | c |} 
 \hline
 $ab\leftrightarrow cd$ & $\sum_{bcd}Q^{ab}_{cd}/g^4 = \sum_{bcd}1/\nu_a|\mathcal{M}^{ab}_{cd}|^2\left[n_b(1\pm n_d)\right]/g^4$ \\ 
 \hline
 $Gg \leftrightarrow Gg$ & $4C_A^2\frac{s^2+u^2}{t^2}n_B(k)\left[1+n_B(k+\omega)\right]+\mathcal{O}(\frac{T^2}{p^2})$ \\ 
 \hline
 $Gq\leftrightarrow Gq$ & $2N_f\cdot4\frac{d_F}{d_A}C_FC_A \frac{s^2+u^2}{t^2}n_F(k)\left[1-n_F(k+\omega)\right]+\mathcal{O}(\frac{T^2}{p^2})$\\ 
 \hline
 $Qq\leftrightarrow Qq$ & $2N_f\cdot4\frac{d_F}{d_A}C_F^2\frac{s^2+u^2}{t^2}n_F(k)\left[1-n_F(k+\omega)\right]+\mathcal{O}(\frac{T^2}{p^2})$ \\
 \hline
 $Qg \leftrightarrow Qg$ & $4C_FC_A\frac{s^2+u^2}{t^2}n_B(k)\left[1+n_B(k+\omega)\right]+\mathcal{O}(\frac{T^2}{p^2})$ \\
 \hline
 $Gq \leftrightarrow Qg$ & $2N_f \cdot 4\frac{d_F}{d_A}C_F^2\frac{u}{t}n_F(k)\left[1+n_B(k+\omega)\right]+\mathcal{O}(\frac{T^2}{p^2})$ \\
 \hline
 $Qg \leftrightarrow Gq$ & $4C_F^2\frac{u}{t}n_B(k)\left[1-n_F(k+\omega)\right]+\mathcal{O}(\frac{T^2}{p^2})$\\
 \hline
 $Gg \leftrightarrow Q\bar{q}$ & $2N_f \cdot4C_F^2\frac{u}{t}n_B(k)\left[1-n_F(k+\omega)\right]+\mathcal{O}(\frac{T^2}{p^2})$ \\
 \hline
 $Q\bar{q} \leftrightarrow Gg$ & $4C_F^2\frac{u}{t}n_F(k)\left[1+n_B(k+\omega)\right]+\mathcal{O}(\frac{T^2}{p^2})$ \\
 \hline
\end{tabular}
 \caption{\label{tab:matrix_elements}In this table, we use capital letter $G$ and $Q$ to denote hard gluons and quarks ($p > p_{\textrm{cut}}$), and lowercase letter $g$ and $q$ to denote soft gluons and quarks($p < p_{\textrm{cut}}$). To simplify the notation, we do not specify the quark species. $Q$ and $q$ include the conditions of various quark species, and can also be anti-quark. }
 \end{table}
\end{center}
\egroup
Up to order $T/p$, we have the following kinematics:
\begin{equation}
\begin{split}
    t &= -(-Q)^2=-(P'-P)^2=-\tilde{q}_\perp^2, \\
    s &= -(P+K)^2 \\
      &= \frac{-t}{2q^2}\big[(p+p')(k+k')+q^2 
       -\cos(\phi)\sqrt{(4pp'+t)(4kk'+t)}\big] \\
      &\simeq (2p)\frac{-t}{2q^2}\left[(k+k')-\cos \phi \sqrt{4kk'+t}\right]\left(1+\frac{T}{p}\right), \\
    u &= -(K'-P)^2 = -t-s \\
      &\simeq -s.
\end{split}
\end{equation}
\end{widetext}

\section{Soft conversion process}
\label{app:soft_conv}
Soft conversion is a process where the identity of the hard parton is changed by its interaction with the medium. Diffusion processes only include the identity preserving soft interactions; a soft conversion process is necessary to consider identity non-preserving soft interactions. 

The collision kernel of the soft conversion reads 
\begin{equation}
\begin{split}
    &\mathcal{C}^{2\leftrightarrow 2}_{\textrm{conv}, q_i}[\delta f] = \delta f^{q_i}(\textbf{p})\Gamma_{q\rightarrow g}^\textrm{conv}(p) - \delta f^{g}(\textbf{p})\frac{d_A}{d_F}\Gamma_{g\rightarrow q}^{\textrm{conv}}(p)\\ 
    &\mathcal{C}^{2\leftrightarrow 2}_{\textrm{conv}, \bar{q}_i}[\delta f] = \delta f^{\bar{q}_i}(\textbf{p})\Gamma_{q\rightarrow g}^\textrm{conv}(p) - \delta f^{g}(\textbf{p})\frac{d_A}{d_F}\Gamma_{g\rightarrow q}^{\textrm{conv}}(p)\\ 
    &\mathcal{C}^{2\leftrightarrow 2}_{\textrm{conv}, g}[\delta f] = \sum_{i=1}^{N_f}\left\{\delta f^g(\textbf{p}) \left[\Gamma_{g\rightarrow q_i}^{\textrm{conv}}(p)+\Gamma_{g\rightarrow \bar{q}_i}^{\textrm{conv}}(p)\right]\right. \\
    &-\frac{d_F}{d_A}\left[\delta f^{q_i}(\textbf{p})\Gamma_{q\rightarrow g}^{\textrm{conv}}(p)+\delta f^{\bar{q}_i}(\textbf{p})\Gamma_{\bar{q}\rightarrow g}^{\textrm{conv}}(p)\right]\left.\right\}
\end{split}
\label{eq:cconv}
\end{equation}

As derived in in Section 3.3 of Ref.\cite{Ghiglieri:2015ala}, at leading order, the parton identity exchange rate is
\begin{equation}
    \begin{split}
       \Gamma_{q\rightarrow g}^{\textrm{conv}}(p) =& \frac{g^2C_F}{4p}\int^{\mu_{\tilde q_\perp}} \frac{d^2 q_\perp}{(2\pi)^2}\frac{m_\infty^2}{q_\perp^2+m_\infty^2},   \\
       =& \frac{g^2C_Fm_\infty^2}{16\pi p}\ln\left[1 + \left(\frac{\tilde \mu_{q_\perp}^2}{m_\infty}\right)^2 \right], \\
       \Gamma_{g\rightarrow q}^{\textrm{conv}}(p) =& \frac{d_F}{d_A}\Gamma_{q\rightarrow g}^{\textrm{conv}}(p), 
    \end{split}
    \label{eq:conversion_rate}
\end{equation}
where $m_\infty^2 \equiv g^2C_FT^2/4$ is the asymptotic mass of quarks. 

Given that the rate of these identity non-preserving soft interactions is suppressed by $T/p$ and the energy exchange $\omega$ is small, we neglect the energy loss due to these soft conversion process, and only incorporate the identity exchange. 

In the numerical implementation, at each time step, we change the identity of the leading parton according to the conversion rates in Eqs. \ref{eq:cconv} and \ref{eq:conversion_rate}. 

\section{Splitting approximation process}
\label{app:splittingapprox}

As discussed in the body of the text,
the collision kernel for $2\leftrightarrow 2$ scattering processes can be  simplified when the energy transfer is large.\footnote{We thank Jacopo Ghiglieri for sharing notes on this, which served as the basis for this appendix.}

For simplicity, we will begin the discussion with the pure glue theory. As we will show here, and as is obvious pictorially, the $2 \leftrightarrow 2$ scattering rate with large $\omega$ can be 
written as an effective $1\rightarrow 2$ rate, which takes the form 
\begin{equation}
C^{2\leftrightarrow2}_{\rm split}(\Lambda) 
=  \frac{1}{2} \int^{p - \Lambda}_{\Lambda}  d\omega \, \frac{d\Gamma(p,\omega)}{d\omega} \, ,
\end{equation}
where 
\begin{multline}
   \label{eq:dgammapureglue}
   \frac{d\Gamma(p,\omega)}{d\omega} = \frac{g^4}{8\pi p^3} 
   \frac{P_{gg}^{g}(z) }{  z^2 (1 - z)^2 } \times  \\
   \frac{C_A}{2} \left( 1 -z + z^2 \right) 
   \int \frac{d^2q_\perp}{(2\pi)^2 }   \frac{\hat q(\delta E)}{\delta E^2 } \, ,
\end{multline}
Here we have defined 
\begin{equation}
\delta E \equiv  \frac{pq_{\perp}^2 }{2 p' k' } \, ,
\end{equation}
and for comparison with other litterature we have defined $\hat q(\delta E)$ for the pure glue case~\cite{Ghiglieri:2015ala}
\begin{equation}
\frac{ \hat q(\delta E) }{\delta E^2} \equiv \int \frac{d^3 k}{(2\pi)^3 k} n_B(k) \, 2\pi \delta(k^- - \delta E) \, .
\end{equation}

This is an approximation of the (unscreened) scattering rate given in \eqref{eq:2to2} with the matrix element for the $gg \leftrightarrow gg$ collisions given by
\begin{equation}
|\mathcal M|^2/g^4 = 16 d_A C_A^2 
\left(3 -\frac{su}{t^2} - \frac{st}{u^2} - \frac{tu}{s^2} \right) \, .
\end{equation}
In this kinematic regime, we can neglect the population factors $n^{c}(p')$ and $n^d(k')$. We will  write 
\begin{equation}
\int_{\k} \equiv  \int \frac{d^3k}{(2\pi)^3 2 k}  = \int_K 2\pi \delta_+(K^2) ,
\end{equation}
for the $k$, $p'$, and $k'$ integrals, with $\int_K = \int d^4K/(2\pi)^4$ and $\delta_+(K^2) = \theta(k^0) \delta(K^2)$. Next, we change variables to integrate over 
$Q = P - P'$ instead of $P'$, and use the four momentum constraint to eliminate $K' = K+ Q$, yielding the phase space integral
\begin{align}
C^{2\leftrightarrow2}_{\rm split}(\Lambda) 
=& \frac{1}{4 p \nu_g} \int_{Q,K}  2\pi \delta_+(K^2) \, 2\pi\delta(-2P\cdot Q + Q^2)  \nonumber \\
 & \qquad \times  2\pi \delta(2K \cdot Q + Q^2)   |\M|^2 \delta f(p)  n(k)  \, .
\end{align}
To understand the kinematics of the process, it is convenient to use the light
cone coordinates
where $q^+ = -q_{-} = (q^0 + q^z)/2$ and   $q^-  =q^0 - q^z$  and we take ${\bm p}$ along the 
$z$ direction.  
\begin{align}
   p^+ + k^+ =& \kb^+ + \pb^+, \\
     k^- =&  \kb^-  + \pb^-, \\
   {\bm \k}_\perp =& {\bm \kb}_\perp + {\bm \pb}_\perp,
\end{align}
while the outgoing onshell constraints read, 
\begin{align}
   2 \pb^+ \pb^- +  \pb_\perp^2  =& 0, \\ 
   2 \kb^+ \kb^-  + \kb_\perp^2  =& 0, \
\end{align}
Now, all four components of the momentum $k$ are of order $ \sim T$.  
In order to satisfy the  onshell constraints and energy-momentum conservation, we have  the following scalings with the 
energy of the probe  for the light cone momenta
\begin{align}
 \kb^+  \sim& p, \\
\kb_\perp  \sim &  \sqrt{pT}, \\
 \kb^- \sim&   T.
\end{align}
Thus, the incoming transverse momentum $\bm{k}_\perp \sim T$ can be ignored,  and  transverse momentum conservation fixes that
\begin{equation}
-{\bm \pb}_\perp = {\bm \kb}_\perp =  {\bm q}_\perp.
\end{equation}
Plus-coordinate momentum conservation yields
\begin{equation}
p =  \kb  + p',  \qquad q^{+} = k' = \omega,
\end{equation}
 Minus-coordinate momentum  conservation yields
\begin{align}
   k^{-} =& \frac{p q_\perp^2 }{2 p' k'},    \\
   \pb^{-} =&\frac{q_\perp^2 }{2 p'}  = -q^{-}, \\
   \kb^{-} =&\frac{q_\perp^2 }{2 k'},  
\end{align}
The invariants are
$t = -Q^2$, $s = - 2 P\cdot K $, $u =  2 K' \cdot P$
\begin{subequations}
   \label{eq:stuwapprox}
\begin{align}
   s =& \frac{p^2}{p'k'} q_\perp^2,  \\
   t =&- \frac{p}{p'} q_\perp^2, \\
   u =&- \frac{p}{k'} q_\perp^2,
\end{align}
\end{subequations}
and satisfy $s + t + u=0$.
Now we write 
\begin{equation}
\int_Q = \int \frac{dq^+ dq^-}{ (2\pi)^2} \frac{d^2q_\perp}{(2\pi)^2} \, ,
\end{equation}
and  integrate over $q^-$
\begin{multline}
\int \frac{dq^-}{(2\pi)} 2\pi \delta(-2 P \cdot Q + Q^2)  2\pi \delta(2 K \cdot Q + Q^2)  \\
= \frac{2\pi}{4 \kb \pb} \delta(k^- - \delta E) \, .
\end{multline}

Assembling the ingredients we have
\begin{equation}
C_{\rm split}^{2\leftrightarrow 2}(\Lambda)  = 
\frac{1}{2} \int_\Lambda^{p-\Lambda} d\omega \frac{d\Gamma}{d\omega} \, ,
\end{equation}
where 
\begin{align}
   2\pi  \frac{d\Gamma}{d\omega}
=& \frac{1}{p^3} \frac{| \M|^2/(16 \nu_g) }{ z (1 -z) } \\ 
 & \times \int \frac{d^2q_\perp}{(2\pi)^2}   \int \frac{d^3k}{(2\pi)^3 k }  n_B(k) 2\pi \delta(k^- - \delta E) \, .
\end{align}
where 
\begin{equation}
\frac{|\M|^2 }{16 \nu_g \,z (1-z)} 
\simeq  g^4 C_A^2 \frac{ (1- z + z^2)^2 }{2 z^3 (1 -z)^3 } \, ,
\end{equation}
Reorganizing terms one finds
\begin{align}
  \frac{d\Gamma}{d\omega}
   =& \frac{g^4}{8\pi p^3} \frac{ P^{g}_{gg}(z)   }{ z^2 (1-z)^2 }  \nonumber \\
    & \times  \left( (C_A - \tfrac{1}{2} C_A) z^2  + \tfrac{1}{2} C_A (1 + (1-z)^2 ) \right)   \nonumber \\ 
    & \times \int \frac{d^2q_\perp}{(2\pi)^2 } 
   \int \frac{d^3 k}{(2\pi)^3 k} n_B(k) \, 2\pi \delta(k^- - \delta E) \, ,
\end{align}
in agreement with \eqref{eq:dgammapureglue}.

The analysis can be extended to include quarks.  
Our starting point is again \eqref{eq:2to2}.
As for the pure glue case it is our interest to describe the splitting process where $p'$ and $k'$ are both large. Then we  have as before
\begin{align}
C^{2\leftrightarrow2}_{\rm split}(\Lambda) 
=& \frac{1}{4 p \nu_a}  \sum_{bcd} \int_{Q,K}  2\pi \delta_+(K^2) \, 2\pi\delta(-2P\cdot Q + Q^2)  \nonumber \\
 & \qquad \times  2\pi \delta(2K \cdot Q + Q^2)   |\M^{ab}_{cd}|^2 \delta f^a(p)  n^b(k)  \, .
\end{align}
Now we distinguish two cases: (i) when a gluon is absorbed from the bath,  and (ii) when a quark is absorbed from a bath.

In the first case the gluon is absorbed from the bath and the hard particle splits into flavors $cd$. The differential rate  takes the form 
\begin{multline}
\frac{d\Gamma^{a(g)}_{cd}}{d\omega}
  = \frac{g^4 G^{a}_{cd}(z)}{32\pi p^3} 
   \int \frac{d^2q_\perp}{(2\pi)^2 } \int \frac{d^3 k}{(2\pi)^3k} \\ \times n_B(k) \, 2\pi \delta(k^- - \delta E) \, .
\end{multline}
where the effective splitting rate  are
the matrix elements (see  Table II of \cite{Arnold:2002zm}) 
evaluated using the kinematic approximations of \eqref{eq:stuwapprox}.
\begin{equation}
G^{a}_{cd}(z) \equiv \frac{|\M^{ag}_{cd}|^2/g^4 }{\nu_a z (1 - z)} \, .
\end{equation}
The effective splitting function is for gluon absorption is
\begin{subequations}
\begin{align}
   G^{q}_{qg} =& \frac{ 4 P^q_{qg}(z) }{z^2 (1 - z)^2 }
	\left[\left(\cf-\frac{\ca}{2}\right)z^2
	+\frac{\ca}{2}\left(1+(1-z)^2\right)\right], \\
   G^{g}_{q\bar q}=&	\frac{4 P^g_{q\bar q}(z)}{z^2(1-z)^2}\left[\left(\cf-\frac{\ca}{2}\right)+\frac{\ca}{2} 
 \left(z^2+(1-z)^2\right)\right],  \\ 
 G^{g}_{gg} =& \frac{4 P^g_{gg}(z) }{z^2 (1 -z)^2 } \left[\left(\ca-\frac{\ca}{2}\right)z^2 +\frac{\ca}{2}\left(1+(1-z)^2\right)\right].
\end{align}
\end{subequations}
Here for the process $a \rightarrow cd$ the momentum fraction $z$ is associated with particle $d$, i.e.  $z=k'/p=-q_\perp^2/u$ and $1-z=p'/p=-q_\perp^2/t$.

To find the total rate we must perform the integral over $\omega$. The 
integration is straightforward and yields for gluon absorption
\begin{equation}
\Gamma^{a (g)}_{bc}  = \frac{g^4}{32\pi p} \left(\frac{T^2}{12}\right) \int_{\Lambda/p}^{1-\Lambda/p} dz \, z (1- z) \, G^{a}_{bc}(z) \, .
\end{equation}
The total rate for the splitting process through gluon absorption
$ \Gamma^a(g) = \tfrac{1}{2} \sum_{bc} \ \Gamma^{a (g)}_{bc} $, 
where the factor of $1/2$ is a symmetry factor.  In practice
this symmetry factor is handled by summing over only distinct processes, and, if the final state involves identical particles, by integrating over the distinct phase-space.
In writing this expression we have used the thermodynamic integral, $\int_0^{\infty} dk\, k n_B(k)  = \pi^2 T^2/6$.

The last remaining
integral over $z$ can be done and total rate for gluon absorption takes the form
\begin{equation}
\label{eq:totalglue_absorb}
\Gamma^{a (g)}_{b c} =  \frac{g^4 T^2} {96\pi p }  
\left[ \frac{c_\Lambda}{z_0} +  c_p -  c_{\rm ln} \log(z_0) \right]  \, ,
\end{equation}
where $z_0 =\Lambda/p$. The coefficients, $c_\Lambda$, $c_p$, $c_{\rm ln}$ are in tabular form as in Table~\ref{tab:gluon_absorb}.
\begin{table}
   \begin{tabular}{| L | L | L | L |} \hline
      \Gamma^{a(g)}_{bc}  & c_{\Lambda}  & c_p &  c_{\rm ln} \\ \hline 
      q \rightarrow q g  & 2 \cf \ca  & -\cf\ca + \cf^2/2 &  -2 \cf \ca + \cf^2  \\ 
      \sum_q g \rightarrow q \bar q & 0  &  -N_f \left(\ca/3 +\cf \right)  & N_f \cf \\ g \rightarrow gg  & 4 \ca^2  &  \tfrac{10}{6} \ca^2  &  -4 \ca^2   \\ \hline
   \end{tabular}
   \caption{The total rates for gluon absorption. \label{tab:gluon_absorb}}
\end{table}
We note (again) that the total rate for $g\rightarrow gg$ is $\Gamma^{g(g)}_{gg}/2$ to account for the symmetry of the final state. 
We also note that the second row in this table has been summed over quark flavors.

We will now consider the case when a soft quark is absorbed from the bath,  and the hard particle of type $a$ splits $a \rightarrow cd$. The differential rate now takes the form
\begin{equation}
\frac{d\Gamma^{a(q)}_{cd}}{d\omega}
  = \frac{g^4 F^{a}_{cd}(z)}{32\pi p^3}  
    \int \frac{d^2q_\perp}{(2\pi)^2 } \int \frac{d^3 k}{(2\pi)^3 k} n_F(k) \, 2\pi \delta(k^- - \delta E) \, .
\end{equation}
where
\begin{equation}
F^{a}_{cd}(z) \equiv \frac{|\M^{aq}_{cd}|^2/g^4 }{\nu_a z (1 - z)} \, .
\end{equation}
Evaluating the matrix elements (again using Table II. of \cite{Arnold:2002zm} and \eqref{eq:stuwapprox}), we find
\begin{widetext}
   \begin{subequations}
\begin{align}
   F^{q_1}_{q_1q_2} =& \frac{2C_F}{z (1-z)} 
   \left[ \frac{1 + (1-z)^2}{z^2} \right],  \\
   F^{q_1}_{q_1q_1} =& \frac{2C_F}{z (1-z)}  \Big[
       \frac{1 + (1-z)^2}{z^2}  + \frac{1 + z^2}{(1-z)^2 } %
    + 4 \left(C_F - \frac{C_A}{2} \right) \frac{1}{z (1-z) } \Big], \\
   F^{q_1}_{q_1\bar q_1} =& \frac{2C_F}{z (1-z)}  \Big[
      \frac{1 + (1-z)^2}{z^2}  + z^2 + (1-z)^2 %
   - 4 \left(C_F - \frac{C_A}{2} \right) \frac{(1-z)^2}{z} \Big],\\
   F^{q_1}_{q_2\bar q_2} =& \frac{2C_F}{z (1-z)}  \left[
   z^2 + (1-z)^2 \right], \\
   F^{q_1}_{gg} =& 4C_F \frac{z^2 + (1-z)^2}{z^2 (1-z)^2}  \Big[
   \left(C_F - \frac{C_A}{2} \right)%
   + \frac{C_A}{2} (z^2 + (1- z)^2 ) 
    \Big], \\
    F^{g}_{q_1 g}=&\frac{4 d_F C_F}{d_A} \frac{1 + z^2}{z^2 (1-z)^3}  \Big[
    \left(C_F - \frac{C_A}{2}\right)(1-z)^2%
 + \frac{C_A}{2}(1+z^2)\Big].
\end{align}
\end{subequations}
Again integrating over the momentum fraction we find that the total rate
takes the form
\begin{equation}
\Gamma^{a(q)}_{cd} =  \frac{g^4} {32\pi p } \left(\frac{T^2}{24}\right)
\left[ \frac{c_\Lambda}{z_0} +  c_p -  c_{\rm ln} \log(z_0) \right]  \, ,
\end{equation}
where we used the integral, $\int_0^{\infty}dp \, p \, n_F(p) = \pi^2 T^2/12$.
The coefficients $c_\Lambda$, $c_p$ and $c_{\rm ln}$ are tabulated in Table \ref{tab:quark_absorb}. 
   \begin{center}
   \begin{table}[h!]
   \begin{tabular}{|L |L | L | L|} \hline
      \Gamma^{a(q)}_{bc}  & c_{\Lambda}  & c_p &  c_{\rm ln} \\ \hline 
      \sum_{q_2} \, q_1 \rightarrow q_1 q_2 & & &  \\ \qquad + q_1 \rightarrow q_1 \bar{q}_2  \,  & 4 \cf (2N_f-2)  & -2\cf (2N_f -2) &  -4\cf  (2N_f-2)  \\ 
      q_1 \rightarrow q_1 q_1  & 8 \cf  & -4\cf &  -8 \cf(1 + \ca -2\cf)   \\ 
      q_1 \rightarrow q_1 \bar q_1  & 4 \cf  & \tfrac{2}{3} \cf(-1 -9\ca +18\cf)  &  -4\cf (1 - \ca + 2\cf)  \\
      \sum_{q_2} q_1 \rightarrow q_2 \bar q_2  & 0 & \tfrac{4}{3} \cf (N_f-1) & 0  \\
      q_1 \rightarrow gg  & 0 &  - \tfrac{8\cf}{3} (\ca + 3\cf) & 8\cf^2  \\
      \sum_{q_1} g \rightarrow q_1g  & & & \\
       \qquad + g \rightarrow \bar{q}_1 g & 4 \ca (2N_f) &  (\cf -2 \ca) (2N_f) & (2\cf - 4 \ca ) (2N_f)  \\
      \hline
   \end{tabular}
   \caption{The total rates for quark absorption. \label{tab:quark_absorb}}
   \end{table}
\end{center}
\end{widetext}

The full transition  rate for collisional-splittings takes the form
\begin{equation}
C^{2 \leftrightarrow 2}_{\rm split}(\Lambda) =  \frac{1}{2} \sum_{cd} \int_\Lambda^{p-\Lambda} d\omega \,  \left( \frac{d{\Gamma}^{a(g)}_{cd} }{d \omega}  + \frac{d\Gamma^{a(q)}_{cd}}{d\omega} \right) \, , 
\end{equation}
and includes  both the gluon and quark induced splittings.

\section{Detailed balance of the Langevin model}

\label{app:detailed_balance}

\begin{figure}
    \centering
    \includegraphics[width=\linewidth]{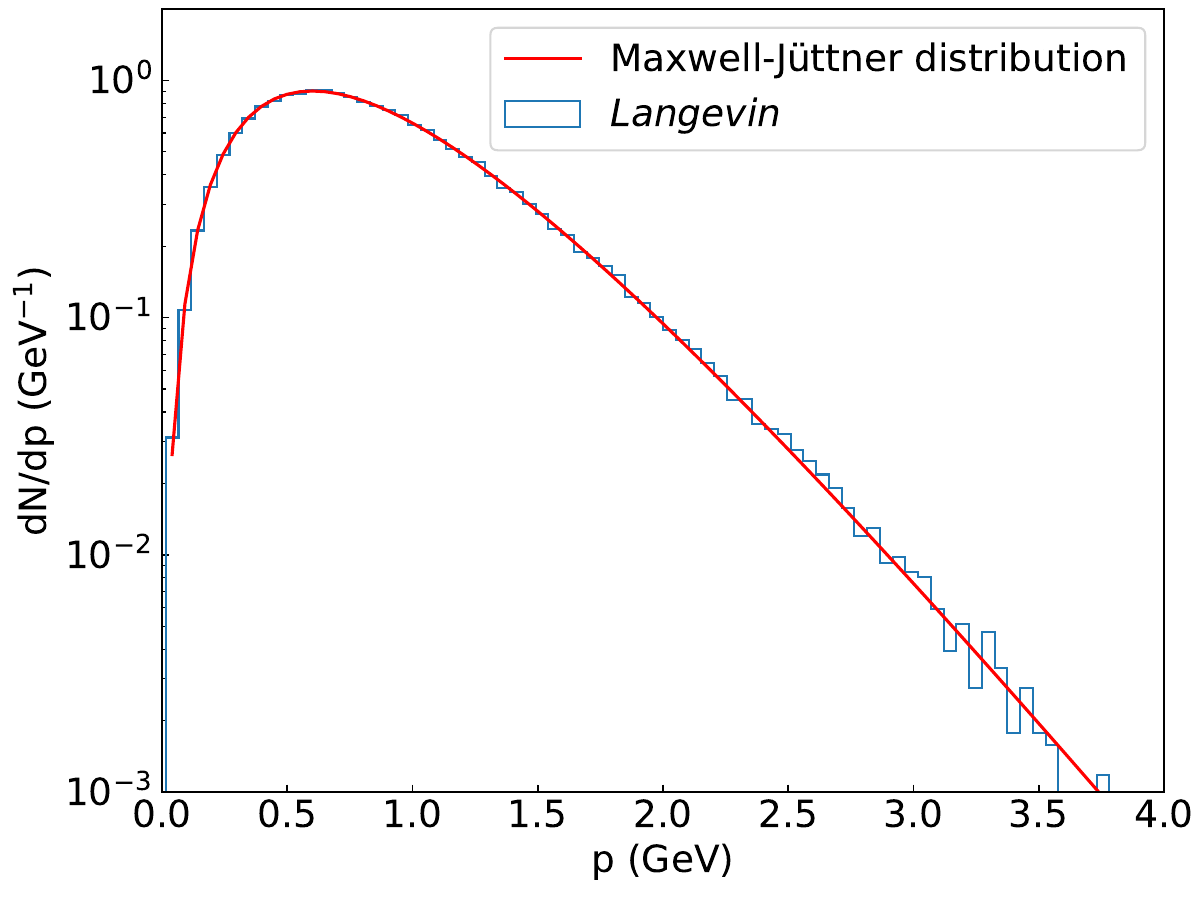}
    \caption{Distribution of momentum from a gluon with initial energy $E_0 = 16$~GeV evolved for a time $t=100$~fm in a $300$~MeV static medium, compared with the thermal distribution. We used $N_f=3$ and $\alpha_s=0.3$. }
    \label{fig:equilibrium}
\end{figure}

\begin{figure*}[tbhp]
    \centering
    \includegraphics[width=\linewidth]{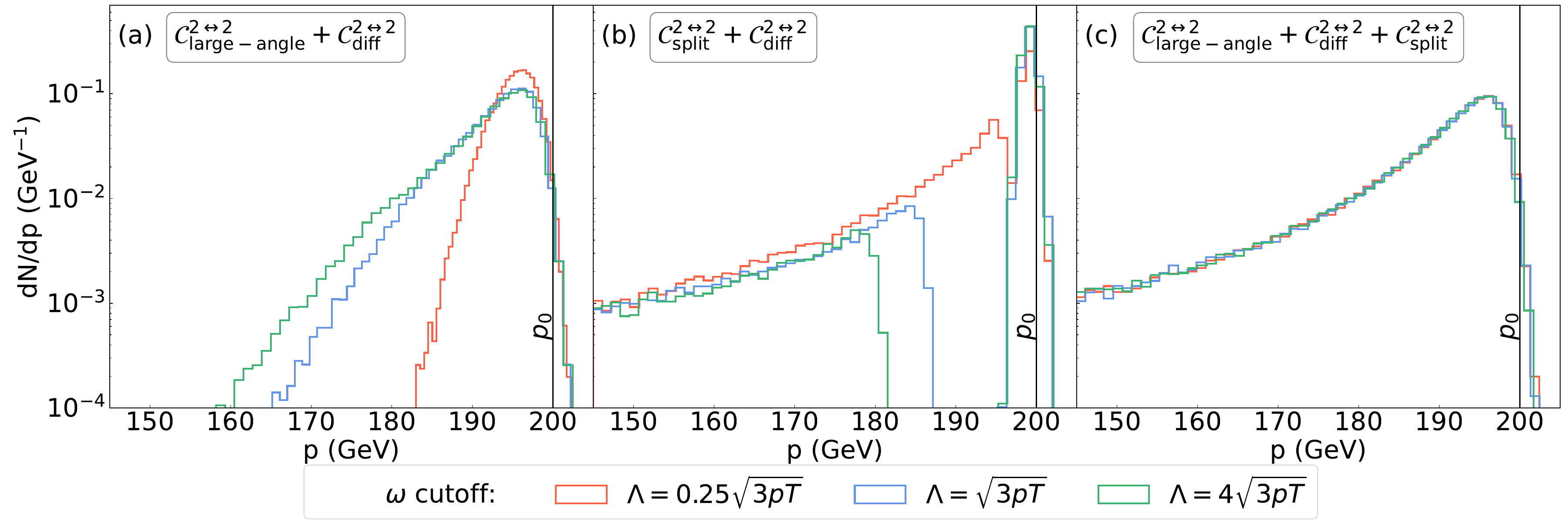}
    \caption{Momentum distributions of final gluons from an initial  gluon with energy $E_0 = 200$~GeV after $1\,{\rm fm}$ of evolution in a $300$ MeV static medium. The system is evolved with elastic $2\leftrightarrow 2$ interactions only, for three different prescriptions for the cutoff $\Lambda$: $\Lambda =\{0.25,1,4\}\min(\sqrt{3 p T},\pcut)$.  The three panels show how the $2\leftrightarrow2$ processes is divided into subprocesses: (a) a large-angle process with soft drag\&diffusion; (b) a splitting process with soft drag\&diffusion; and (c) the full $2\leftrightarrow2$ rate including large-angle scattering, splitting, and soft drag\&diffusion. The full rate shown in (c) is approximately independent of the prescription for $\Lambda$.}
    \label{fig:Lambda}
\end{figure*}

The diffusion process as described by the Fokker-Planck equation (\eqref{eq:FP}) can be stochastically realized with the Langevin model. The stochastic Langevin equations solves the evolution of the space-time coordinates and the momentum of the particle~\cite{Dunkel:2009tla, He:2013zua}: 
\begin{equation}
    \begin{split}
        \frac{\Delta\bm{x}}{\Delta t} &= \frac{\bm{p}}{E}\\
        \frac{\Delta\bm{p}}{\Delta t} &= -\eta_{D, \textrm{soft}}\bm{p}+\bm{F}^{\textrm{thermal}}(t)
    \end{split}
    ,
\end{equation}
where $\bm{x}$ is the space coordinates of the parton, $\bm{F}^{\textrm{thermal}}$ is a thermal random force satisfying the mean and the correlation function 
\begin{equation}
\begin{split}
    \left<F^{\textrm{thermal}}_i\right> &= 0 \\
    \left<F^{\textrm{thermal}}_i F^{\textrm{thermal}}_j\right> &= -\frac{1}{\Delta t}\left[\hat{p}_i\hat{p}_j\hat{q}_L+\frac{1}{2}\left(\delta_{ij}-\hat{p}_i\hat{p}_j\right)\hat{q}\right]
\end{split}. 
\label{eq:Langevin}
\end{equation}
The realization of the stochastic differential equation is dependent on the discretization scheme. We choose the pre-point Ito scheme in this work~\cite{Moore:2004tg}. In the infinite medium limit, the initial energetic partons should eventually reach the thermal equilibrium via diffusion in the thermal plasma. The equilibrium distribution of the light parton $\delta f(p)$ is proportional to $\exp (-p/T)$ in the Fokker-Planck equation (\eqref{eq:FP}), and the time derivative of the equilibrium distribution is zero. We can thus obtain the drag coefficient $\eta_{D, \textrm{soft}}$ as \eqref{eq:drag}.

We check the thermalization of the light partons in the QGP plasma using the Langevin model (\eqref{eq:Langevin}) with the drag and diffusion coefficients in Equation (\ref{eq:inel_qhat_L}-\ref{eq:drag}). 
As shown in Figure~\ref{fig:equilibrium}, after a long evolution time, the momentum distribution of the light parton approaches the Maxwell-J\"uttner distribution~\cite{juttner1911maxwellsche}\cite{[{See also }][{, and references therein.}]Mendoza:2012tr}:

\begin{equation}
    \delta f(p) \propto p^2\exp \left(-\frac{E}{T}\right). 
    \label{eq:Maxwell-Juttner}
\end{equation}

\section{$\Lambda$ cutoff dependence}

\label{app:lambda_dependence}

As described in Section II, the hard elastic interactions are divided as the large-angle process and the splitting approximation process. In Fig. \ref{fig:Lambda}, we show the evolution of a gluon in quark-gluon plasma ($N_f=3$) with only $2\leftrightarrow 2$ interactions. In Fig. \ref{fig:Lambda}(a), with only $\mathcal{C}^{2\leftrightarrow 2}_{\mathrm{large-angle}}$ and $\mathcal{C}^{2\leftrightarrow 2}_{\mathrm{diff}}$, the tail of the energy distribution depends significantly on the value of $\Lambda$. 
In Fig. \ref{fig:Lambda}(b), with only $\mathcal{C}^{2\leftrightarrow 2}_{\mathrm{diff}}$ and $\mathcal{C}^{2\leftrightarrow 2}_{\mathrm{split}}$, the interactions with $\tilde{q}_\perp > \mu_{\tilde{q}_\perp}$ and $\omega < \Lambda$ is missed, which result in a missing part of the energy distribution; inevitably, the energy distribution around the initial parton energy $p_0$ is found to depend on $\Lambda$. In Fig. \ref{fig:Lambda}(c), with all the types of the $2\leftrightarrow 2$ interactions combined ($\mathcal{C}^{2\leftrightarrow 2}_{\mathrm{large-angle}}$+$\mathcal{C}^{2\leftrightarrow 2}_{\mathrm{diff}}$+$\mathcal{C}^{2\leftrightarrow 2}_{\mathrm{split}}+\mathcal{C}^{2\leftrightarrow 2}_{\mathrm{conv}}$), the result is found to be independent of the cutoff $\Lambda$, as expected.

\section{Propagation of energetic light quarks}
\label{app:quark_energy_distribution}
\begin{figure*}[tbhp]
    \centering
    \includegraphics[width=2\columnwidth]{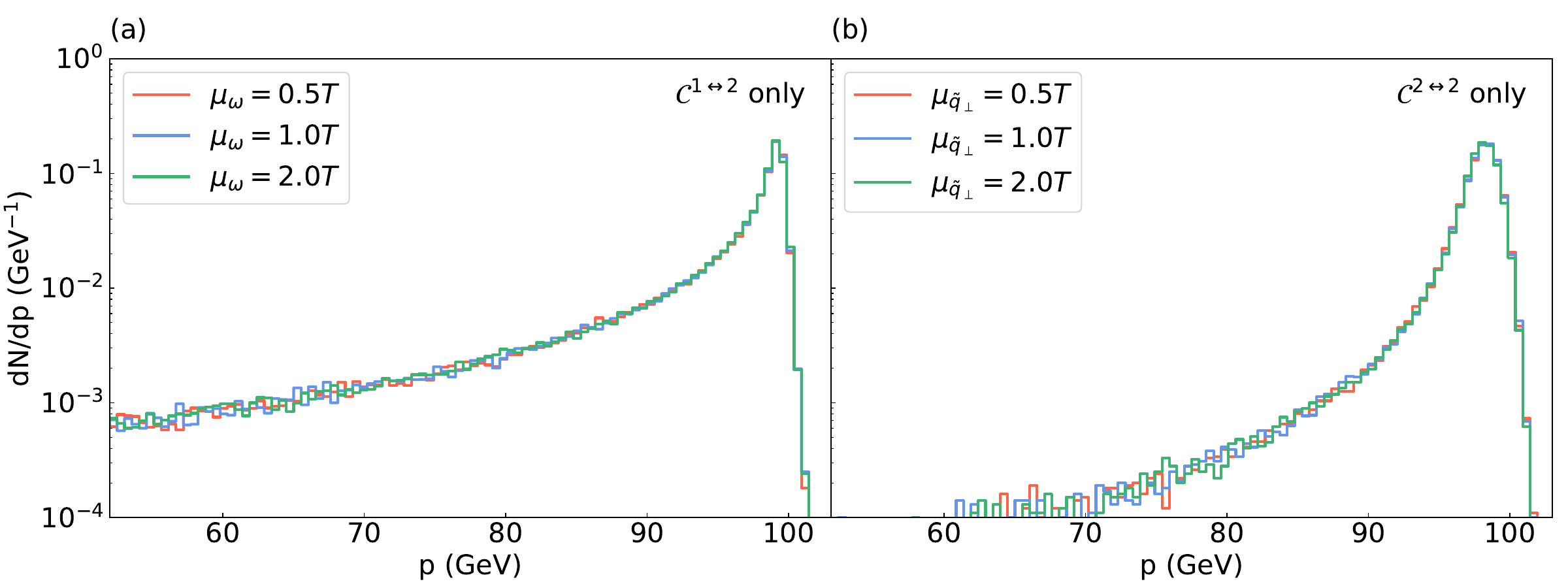}

    \caption{The energy distribution of a $100~\textrm{GeV}$ ``up" quark propagating through $300~\textrm{MeV}$ QGP medium ($N_f=3$) at $\alpha_s = 0.005$ with different values of the cutoff. The evolution time is $t=(0.3/\alpha_s)^2=3600~\textrm{fm}$. The subplot (a)  only includes $\mathcal{C}^{1\leftrightarrow 2}$ interactions and (b) only includes $\mathcal{C}^{2\leftrightarrow 2}$ interactions. }
    \label{fig:cutoff_dependence_small_coupling_quark}
\end{figure*}
\begin{figure*}[tbhp]
    \centering
    \includegraphics[width=2\columnwidth]{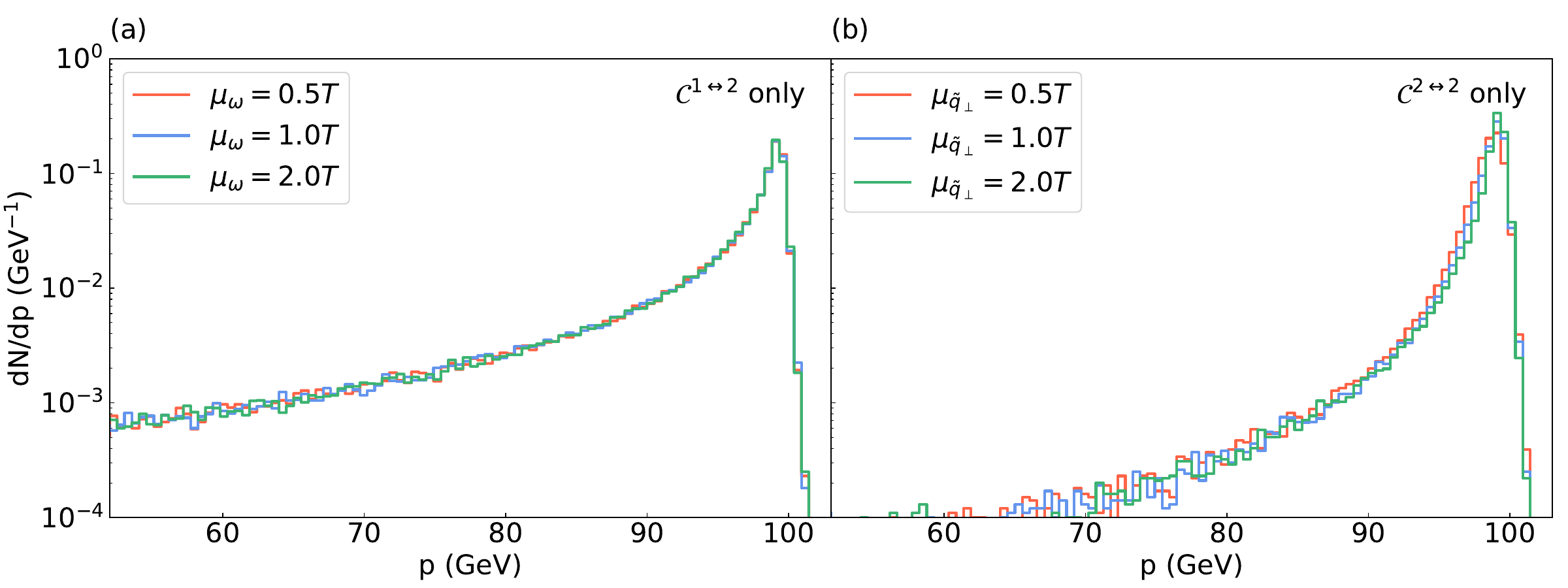}

    \caption{The energy distribution of a $100~\textrm{GeV}$ ``up" quark propagating through $300~\textrm{MeV}$ QGP medium ($N_f=3$) at $\alpha_s = 0.3$ with different values of the cutoff. The evolution time is $t=(0.3/\alpha_s)^2=1~\textrm{fm}$. The subplot (a)  only includes $\mathcal{C}^{1\leftrightarrow 2}$ interactions and (b) only includes $\mathcal{C}^{2\leftrightarrow 2}$ interactions. See the weakly-coupled results in Fig.~\ref{fig:cutoff_dependence_small_coupling_quark} for comparison. }
    \label{fig:cutoff_dependence_large_coupling_quark}
\end{figure*}

In Sections \ref{sec:parton_evolution_small} and \ref{sec:parton_evolution_large}, we presented the propagation of a hard gluon in a static quark-gluon plasma ($N_f=3$) at both small and large coupling. The energy distribution of this hard gluon evolution was presented for different values of hard-soft cutoffs in Figures \ref{fig:cutoff_dependence_small_coupling} and \ref{fig:cutoff_dependence_large_coupling}. In this appendix, we perform the same tests for a hard ``up'' quark: Figures \ref{fig:cutoff_dependence_small_coupling_quark} and \ref{fig:cutoff_dependence_large_coupling_quark}. The conclusion are the same for the evolution of a quark and that of a gluon. In the small coupling regime ($\alpha_s = 0.005$), both $1\leftrightarrow 2$ interactions and $2\leftrightarrow 2$ interactions are independent of the hard-soft cutoff. In the larger coupling regime ($\alpha_s = 0.3$), $1\leftrightarrow 2$ interactions is still independent on the hard-soft cutoff, while for $2\leftrightarrow 2$ interactions, there is a slightly larger cutoff dependence around the initial energy. 

\bibliography{reference}

\end{document}